\documentclass[prb,twocolumn,showpacs,groupedaddress]{revtex4}
\usepackage{amssymb,amsmath}
\usepackage{graphicx,array}
\usepackage{dcolumn}
\usepackage{bm}

\begin{document}
\title{Insights into the Phase Diagram of Bismuth Ferrite 
          from Quasi-Harmonic Free Energy Calculations}

\author{Claudio Cazorla and Jorge \'I\~niguez}
\affiliation{Institut de Ci\`encia de Materials de Barcelona (ICMAB-CSIC),
             Campus UAB, 08193 Bellaterra, Spain} 

\begin{abstract}
We have used first-principles methods to investigate the phase diagram
of multiferroic bismuth ferrite (BiFeO$_{3}$ or BFO), revealing the
energetic and vibrational features that control the occurrence of 
various relevant structures. More precisely, we have studied the
relative stability of four low-energy BFO polymorphs by computing
their free energies within the quasi-harmonic approximation,
introducing a practical scheme that allows us to account for the main
effects of spin disorder. As expected, we find that the ferroelectric
ground state of the material (with $R3c$ space group) transforms into
an orthorhombic paraelectric phase ($Pnma$) upon heating. We show that
this transition is not significantly affected by magnetic disorder, and
that the occurrence of the $Pnma$ structure relies on its being
vibrationally (although not elastically) softer than the $R3c$
phase. We also investigate a representative member of the family of
{\sl nano-twinned} polymorphs recently predicted for BFO [Prosandeev
  {\sl et al}., Adv. Funct. Mater. {\bf 23}, 234 (2013)] and discuss
their possible stabilization at the boundaries separating the $R3c$
and $Pnma$ regions in the corresponding pressure-temperature phase
diagram. Finally, we elucidate the intriguing case of the so-called
{\sl super-tetragonal} phases of BFO: Our results explain why such
structures have never been observed in the bulk material, despite
their being stable polymorphs of very low energy. Quantitative
comparison with experiment is provided whenever possible, and the
relative importance of various physical effects (zero-point motion,
spin fluctuations, thermal expansion) and technical features (employed 
exchange-correlation energy density functional) is discussed. Our work 
attests the validity and usefulness of the quasi-harmonic scheme to 
investigate the phase diagram of this complex oxide, and prospective 
applications are discussed.
\end{abstract}


\maketitle

\section{Introduction}
\label{sec:intro}

Magnetoelectric multiferroics, a class of materials in which
ferroelectric and (anti-)ferromagnetic orders coexist, are generating
a flurry of interest because of their fundamental complexity and
potential for applications in electronics and data-storage devices,
among others. In particular, the magnetoelectric coupling between
their magnetic and electric degrees of freedom opens the possibility
for the control of the magnetization {\sl via} the application of a
bias voltage in advanced spintronic
devices.\cite{fiebig05,fiebig06,ramesh07,eerenstein06,balke12,rovillain10,gajek07}

\begin{figure*}[t]
\centerline{
\includegraphics[width=0.90\linewidth]{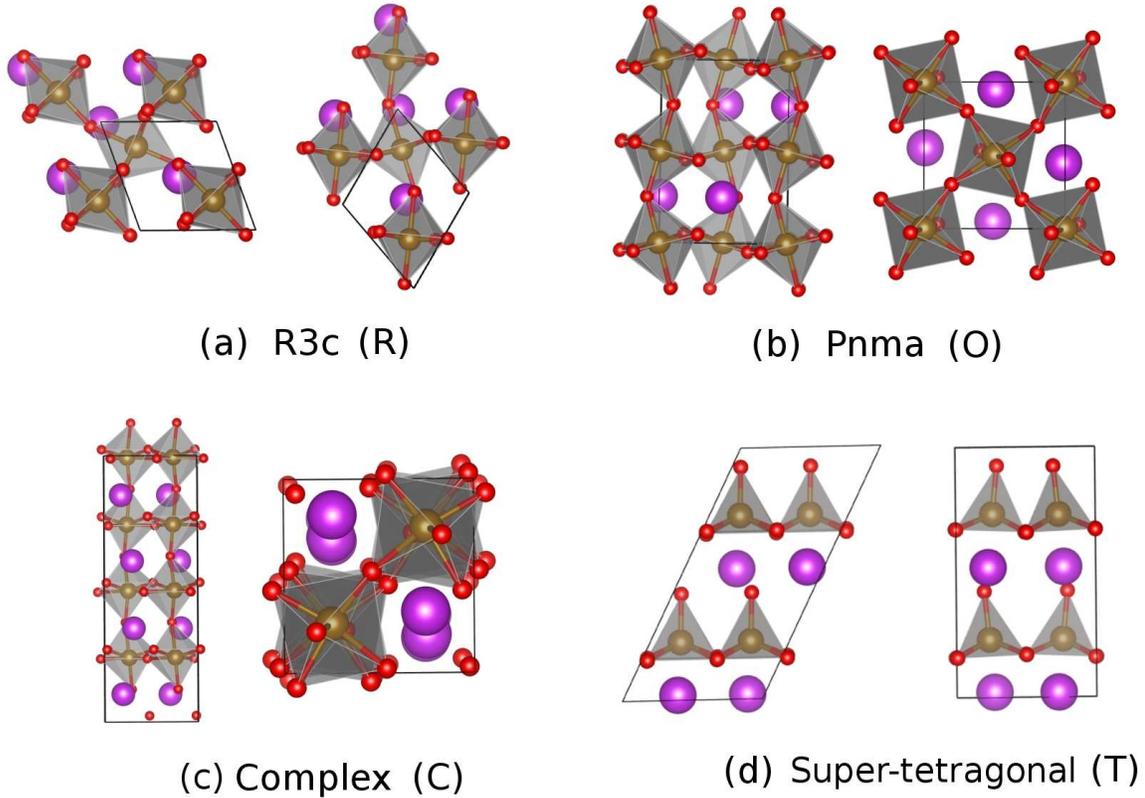}}
\caption{(Color online) Sketch of the four crystal structures considered in this 
  work as seen from two perpendicular directions.  Bi, Fe, and O atoms are
  represented with purple, brown, and red spheres, respectively. Unit
  cells are depicted with thick solid lines and the O$_{6}$ octahedra
  and O$_{5}$ pyramids appear shadowed.}
\label{fig1}
\end{figure*}

Perovskite oxide bismuth ferrite (BiFeO$_{3}$ or BFO) is the
archetypal single-phase multiferroic compound. This material possesses
unusually high anti-ferromagnetic N{\`e}el and ferroelectric Curie
temperatures ($T_{\rm N} \sim 650$~K and $T_{\rm C} \sim 1100$~K,
respectively\cite{kiselev63,smolenskii61,catalan09}) and, remarkably,
room-temperature magnetoelectric coupling has been experimentally
demonstrated in BFO thin films and single
crystals.\cite{lee08,lebeugle08,zhao06} Under ambient conditions, bulk
BFO has a rhombohedrally distorted structure with the $R3c$ space
group [see Fig.~\ref{fig1}(a)]; such a structure can be derived from
the standard cubic {\sl AB}O$_{3}$ perovskite phase by simultaneously
condensing ($i$) a polar cation displacement accompanied by an
unit-cell elongation along the $[111]$ pseudo-cubic direction, and
($ii$) anti-phase rotations of neighboring oxygen octahedra about the
same axis (this is the rotation pattern labeled by $a^{-}a^{-}a^{-}$
in Glazer's notation\cite{glazer}). The basic magnetic structure is
anti-ferromagnetic G-type (G-AFM), so that first-nearest-neighboring
iron spins are anti-aligned; superimposed to this G-AFM arrangement,
in bulk samples there is an incommensurate cycloidal modulation.

Interestingly, in spite of the extensive studies performed, there are
still a few controversial aspects concerning the pressure-temperature
($p-T$) phase diagram of BFO. Under ambient pressure BFO transforms
from the $R3c$ phase to a paramagnetic $\beta$-phase at the Curie
temperature $T_{\rm C} \sim 1100$~K; upon a further temperature
increase of about $100$~K, the compound transforms to a cubic
$\gamma$-phase that rapidly decomposes and melts at about 1250~K. The
exact symmetry of the paramagnetic $\beta$-phase has been contentious
for some time. Based on Raman measurements, Haumont {\sl et
  al}.\cite{haumont06} suggested that this was a cubic
$Pm\overline{3}m$ structure; however, subsequent thermal,
spectroscopic, and diffraction studies by Palai \emph{et
  al.}\cite{palai08} indexed it as orthorhombic $P2mm$. Next, Kornev
\emph{et al.}\cite{kornev07} predicted the appearance of a tetragonal
$I4/mcm$ phase just above $T_{\rm C}$ using first-principles-based
atomistic models. However, further analysis and experimental XRPF
measurements suggested that this phase is actually monoclinic
$P2_{1}/m$.\cite{haumont08} Lastly, Arnold \emph{et
  al.}~\cite{arnold09} performed detailed neutron diffraction
investigations and arrived at the conclusion that the paramagnetic
$\beta$-phase has the orthorhombic $Pnma$ structure that is
characteristic of GdFeO$_{3}$ [$a^{-}a^{-}c^{+}$ rotation pattern in
  Glazer's notation; see Fig.1(b)].\cite{aclaration-pnma}

The pressure-driven sequence of transitions that BFO presents at room
temperature is not fully understood either. The first-principles study
of Ravindran \emph{et al.} predicted a pressure-induced structural
transition of the $R3c \to Pnma$ type to occur at $p \sim
13$~GPa.~\cite{ravindran06} However, a later synchrotron diffraction
and far-infrared spectroscopy study has suggested that BFO undergoes
two phase transitions below $10$~GPa: the first one at $3.5$~GPa from
the rhombohedral $R3c$ to a monoclinic $C2/m$ structure, and the
second one at $10$~GPa to an orthorhombic $Pnma$
phase.\cite{haumont09} Most recently, Guennou {\sl et
  al}.\cite{guennou11} reported X-ray diffraction and Raman
measurements showing that in the range between 4~GPa and 11~GPa (i.e.,
between the stability regions of the $R3c$ and $Pnma$ phases) there
are three, as opposed to one, different stable structures of BFO. The
authors describe such phases as possessing large unit cells and complex
patterns of O$_{6}$-octahedra rotations and Bi-cation displacements.

Interestingly, BFO's phase diagram was recently reexamined
theoretically by Prosandeev {\sl et al}.\cite{prosandeev12}, employing
an atomistic model that captures correctly the first-principles
prediction\cite{dieguez11} that the $R3c$ and $Pnma$ structures are
local energy minima. These authors found that, at ambient pressure,
the $Pnma$ phase is stable at high temperatures, while the $R3c$
structure is the ground state. Additionally, they predicted an
intermediate orthorhombic phase presenting a complex
octahedral-tilting pattern that can be seen as a {\sl bridge} between
the $a^{-}a^{-}a^{-}$ and $a^{-}a^{-}c^{+}$ cases, with the sequence
of O$_{6}$ rotations along one direction displaying a longer
repetition period. In fact, Prosandeev {\sl et al}. found that there
is a whole family of metastable phases that are competitive in this
temperature range and whose rotation pattern can be denoted as
$a^{-}a^{-}c^{q}$,\cite{bellaiche13} where $q$ is a general wave
vector characterizing the non-trivial modulation of the O$_{6}$ tilts
about the third axis. Figure~\ref{fig1}(c) shows one such phase whose
corresponding $q$-vector is $2\pi/a (1/2,1/2,1/4)$, where $a$ is the
pseudo-cubic lattice constant. There are reasons to believe that such
complex phases can also appear under high-$p$ conditions or upon
appropriate chemical substitutions;\cite{prosandeev12} further, they
seem to be the key to understand the lowest-energy structures
predicted for the ferroelectric domain walls of this
material.\cite{dieguez13}

Finally, another family of novel phases was recently discovered in
strongly-compressed BFO thin films.\cite{bea09,zeches09} These
so-called {\sl super-tetragonal} structures can display aspect ratios
$c/a$ approaching 1.30, and are markedly different from the BFO phases
mentioned above. Various theoretical works have found that many of
them can occur,\cite{dupe10,dieguez11} all being metastable energy
minima of the material.\cite{dieguez11} From the collection of
structures reported by Di\'eguez {\sl et al}.,\cite{dieguez11} a
monoclinic $Cc$ phase with a canted polarization of about
1.5~C/m$^{2}$ and anti-ferromagnetic order of $C$-type (i.e., in-plane
neighboring spins anti-align and out-of-plane neighboring spins
align) emerges as a particularly intriguing case [see
  Fig.~\ref{fig1}(d)]. At $T=0$~K this monoclinic phase turns out to
be energetically very competitive with the paraelectric $Pnma$
structure [see Fig.~\ref{fig1}(b)] that we believe becomes stable at
high temperatures and high pressures. However, to the best of our
knowledge, this super-tetragonal phase has never been observed in BFO
bulk samples suggesting that both temperature and pressure tend
to destabilize it in favor of the $Pnma$ structure.

One would like to use accurate first-principles methods to better
understand what controls the relative stability of the different phases
of BFO, and thus what determines its complex and still debated phase
diagram. However, a direct first-principles simulation of such a
complex material at finite temperatures is computationally very
demanding, and not yet feasible. Within the community working on
ferroelectrics like BaTiO$_3$, PbTiO$_3$ and related compounds, such a
difficulty has been overcome by introducing mathematically simple
effective models, with parameters computed from first principles, that
permit statistical simulations and, thus, the investigation of
$T$-driven
phenomena.\cite{zhong94,zhong95,waghmare97,sepliarsky05,shin05,wojdel13}
In particular, as mentioned above, the so-called {\sl
  effective-Hamiltonian} approach has been also applied to 
BFO,\cite{kornev07} and much effort has been devoted to the
construction of reliable models capturing its structural and magnetic
complexity.\cite{prosandeev12,rahmedov12} Yet, as far as we know, we
still do not have models capable of describing all the relevant BFO
structures mentioned above. Further, BFO has proved to be much more
challenging than BaTiO$_3$ or PbTiO$_3$ for model-potential work;
thus, a direct and accurate first-principles treatment is highly
desirable.

Fortunately, BFO presents a peculiar feature that enormously
simplifies the investigation of its phase diagram. Unlike the usual
ferroelectric materials, whose transitions are typically driven by the
condensation of a soft phonon mode, BFO presents strongly first-order
{\sl reconstructive} transformations between phases that are robustly
metastable. This makes it possible to apply to BFO tools that are
well-known for the analysis of solid-solid phase transitions in other
research
fields,\cite{cazorla08,shevlin12,cazorla12,cazorla10,taioli07,cazorla07,cazorla09}
and which are based on the calculation of the free energy of the
individual phases as a function of temperature, pressure, etc. The
simplest of such techniques, which requires relatively affordable
first-principles simulations, is based on a {\sl quasi-harmonic}
approximation to the calculation of the free energy (QHF method in the
following). This is the scheme adopted in this work to investigate
BFO's $p-T$ phase diagram.

We should stress, though, that application of the QHF scheme to BFO is
not completely straightforward. Indeed, the spin and vibrational
degrees of freedom in multiferroic materials can be expected to couple
significantly (i.e., spin-phonon coupling effects become
non-negligible,\cite{fennie06,hee10,hemberger06,rudolf08,hong12} see
Fig.~\ref{fig2}) implying that the free energies of ferromagnetic,
anti-ferromagnetic, and paramagnetic phases belonging to a same
crystal structure may differ significantly. The situation becomes
especially complicated whenever we have structural transitions
involving paramagnetic phases, as capturing the effect of disordered
spin arrangements would in principle require the use of very large
simulation boxes.\cite{kormann08,shang10,kormann12} In this
work we have introduced and applied an approximate scheme to
circumvent such a difficulty.

Therefore, here we present a QHF investigation of the $p-T$ phase
diagram of BFO, monitoring the relative stability of the four
representative phases shown in Fig.~\ref{fig1}: the rhombohedral
ground state (``${\cal R}$ phase'' with $R3c$ space group), the
orthorhombic structure that gets stabilized at high temperatures and
pressures (``${\cal O}$ phase'' with $Pnma$ space group), a phase that
is representative of the recently predicted {\sl nano-twinned}
structures displaying complex O$_6$-rotation patterns (complex or
``${\cal C}$ phase''), and the most stable of the super-tetragonal
polymorphs that have been predicted to occur in strongly-compressed
thin films (``${\cal T}$ phase'' with $Cc$ space group). Our
calculations take into account the fluctuations of spin ordering in an
approximate way and reveal the subtle effects that control the
occurrence (or suppression) of all these structures in BFO's phase
diagram.

The organization of this article is as follows. In Section~II we
provide the technical details of our energy and phonon calculations,
and briefly review the fundamentals of the QFH approach. We also
explain the strategy that we have followed to effectively cast
spin-phonon coupling effects into QHF expressions. In Section~III we
present and discuss our results. Finally, in Section~IV we conclude
the article by reviewing our main findings and commenting on
prospective work.

\section{Methods}
\label{sec:methods}

\subsection{First-principles methods}

In most of our calculations we used the generalized gradient
approximation to density functional theory (DFT) proposed by Perdew,
Burke, and Ernzerhof (GGA-PBE),\cite{pbe96} as implemented in the VASP
package.\cite{vasp} We worked with GGA-PBE because this is the DFT
variant that renders a more accurate description of the relative
stability of the ${\cal R}$ and ${\cal O}$ phases of BFO, as discussed
in Ref.~\onlinecite{dieguez11}. A ``Hubbard-U'' scheme with $U = 4$~eV
was employed for a better treatment of Fe's $3d$ electrons.  We used
the ``projector augmented wave'' method to represent the ionic
cores,\cite{bloch94} considering the following electrons as valence
states: Fe's $3s$, $3p$, $3d$, and $4s$; Bi's $5d$, $6s$, and $6p$;
and O's $2s$ and $2p$. Wave functions were represented in a plane-wave
basis truncated at $500$~eV, and each crystal structure was studied on
its corresponding unit cell (see Fig.~\ref{fig1}).  For integrations
within the Brillouin zone (BZ), we employed $\Gamma$-centered $k$-point
grids whose densities were approximately equivalent to that of a $10
\times 10 \times 10$ mesh for the ideal cubic perovskite 5-atom cell
(e.g., $8 \times 8 \times 8$ in the ${\cal R}$ phase with $Z = 2$, and
$6 \times 6 \times 6$ in the ${\cal O}$ phase with $Z=4$). Using these
parameters we obtained energies that were converged to within
$0.5$~meV per formula unit (f.u.). Geometry relaxations were performed
using a conjugate-gradient algorithm that kept the volume of the unit
cell fixed while permitting variations of its shape, and the imposed
tolerance on the atomic forces was $0.01$~eV$\cdot$\AA$^{-1}$.
Equilibrium volumes were subsequently determined by fitting the sets
of calculated energy points to equations of state. Technical details
of our phonon calculations are provided in Secs.~\ref{subsec:phonons}
and~\ref{subsec:lo-to}.

\subsection{Quasi-harmonic Free Energy Approach}
\label{subsec:Fqh}

In the quasi-harmonic approach, one assumes that the potential energy
of the crystal can be captured by a quadratic expansion around the
equilibrium configuration of the atoms, so that
\begin{equation} 
E_{\rm harm} = E_{\rm eq} + \frac{1}{2}
\sum_{l\kappa\alpha,l'\kappa'\alpha'}
\Phi_{l\kappa\alpha,l'\kappa'\alpha'} u_{l\kappa\alpha}
u_{l'\kappa'\alpha'} \; ,
\label{eq:eqh}
\end{equation}
where $E_{\rm eq}$ is the total energy of the undistorted lattice,
$\boldsymbol{\Phi}$ the force-constant matrix, and $u_{l\kappa\alpha}$
is the displacement along Cartesian direction $\alpha$ of the atom
$\kappa$ at lattice site $l$. In the usual way, we tackle the
associated dynamical problem by introducing
\begin{equation}
u_{l\kappa\alpha}(t) = \sum_{q} u_{q\kappa\alpha} \exp{ \left[ i
    \left(\omega t - \boldsymbol{q} \cdot (\boldsymbol{l}+
    \boldsymbol{\tau}_{\kappa} \right) \right] } \; ,
\end{equation}
where $\boldsymbol{q}$ is a wave vector in the first Brillouin zone (BZ)
defined by the equilibrium unit cell;
$\boldsymbol{l}+\boldsymbol{\tau}_{\kappa}$ is the vector that locates
the atom $\kappa$ at cell $l$ in the equilibrium structure. Then, the
normal modes are found by diagonalizing the dynamical matrix
\begin{equation}
\begin{split}
& D_{\boldsymbol{q};\kappa\alpha,\kappa'\alpha'} =\\ &
  \frac{1}{\sqrt{m_{\kappa}m_{\kappa'}}} \sum_{l'}
  \Phi_{0\kappa\alpha,l'\kappa'\alpha'} \exp{\left[
      i\boldsymbol{q}\cdot(\boldsymbol{\tau}_{\kappa}-\boldsymbol{l'}-\boldsymbol{\tau}_{\kappa'})
      \right]} \, ,
\end{split}
\end{equation}
and thus treat the material as a collection of non-interacting
harmonic oscillators with frequencies $\omega_{\boldsymbol{q}s}$
(positively defined and non-zero) and energy levels
\begin{equation}
E^{n}_{\boldsymbol{q}s} = \left( \frac{1}{2} + n \right)
\omega_{\boldsymbol{q}s} \, ,
\end{equation}
where $0 \le n < \infty$. Within this approximation, the Helmholtz
free energy at volume $V$ and temperature $T$ is given by
\begin{equation}
F_{\rm harm} (V,T) = \frac{1}{N_{q}}~k_{B} T \sum_{{\bf
    q}s}\ln\left[ 2\sinh \left( \frac{\hbar\omega_{{\bf
        q}s}(V)}{2k_{\rm B}T} \right) \right] \; ,
\label{eq:fharm}
\end{equation}
where $N_{q}$ is the total number of wave vectors used in our
BZ integration and the dependence of frequencies
$\omega_{\boldsymbol{q}s}$ on volume is indicated. Finally, the total
Helmholtz free energy of the crystal can be written as
\begin{equation}
F_{\rm qh} (V,T) = E_{\rm eq} (V) + F_{\rm harm} (V,T) \; .
\label{eq:fqh}
\end{equation}   
We note that the greater contributions to $F_{\rm harm}$ come from the
lowest-frequency modes.  This implies that, when analyzing the
thermodynamic stability of different crystal structures, those which
are vibrationally softer in average will benefit more from the
dynamical term in $F_{\rm qh}$.
 
Finally, let us analyze the form that $F_{\rm harm}$ adopts in the
limits of low and high temperatures. In the first case, one obtains
\begin{equation}
F_{\rm harm} (V,T \to 0) = \frac{1}{N_{\rm q}} \sum_{\boldsymbol{q}s}
\frac{1}{2}\hbar\omega_{\boldsymbol{q}s} \; ;
\label{eq:zpe}
\end{equation}
this result is usually referred to as the zero-point energy (ZPE).
As we will see in Sec.~\ref{sec:results}, ZPE corrections
may turn out to be decisive in the prediction of accurate transition
pressures involving two crystal structures with similar static
energies. In the second limiting case, usually termed as the {\sl classical
  limit} (i.e., for $\hbar\omega_{\boldsymbol{q}s} \ll k_{B} T$), one
arrives at the expression
\begin{equation}
F_{\rm harm} (V,T \to \infty) = 3 N_{\rm uc} k_{B} T \ln\left[\frac{\hbar
    \bar{\omega}}{k_{B} T}\right] \; .
\label{eq:classfharm}
\end{equation}
Here, $N_{\rm uc}$ is the number of atoms in the unit cell, and $\bar{\omega}$
is the geometric average frequency defined as
\begin{equation}
\bar{\omega} = \exp{(\langle \ln \omega \rangle)} \; ,  
\label{eq:omegabar}
\end{equation}
where $\langle \cdots \rangle$ is the arithmetic mean performed over
wave-vectors $\boldsymbol{q}$ and phonon branches $s$.  It is worth
noting that low-frequency modes are the ones contributing the most to
$\bar{\omega}$, and therefore to $F_{\rm harm}$.  As it will be shown
in the next section, Eq.~(\ref{eq:classfharm}) allows us to obtain
compact and physically insightful expressions for $F_{\rm harm}$ in
which spin-phonon coupling effects are effectively accounted for.

\begin{figure}
\centerline{
\includegraphics[width=1.00\linewidth]{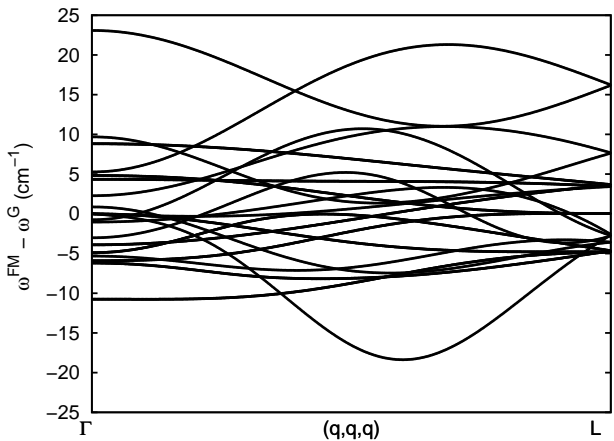}}
\caption{Phonon frequency shifts among the ferromagnetic (FM) and
  G-type anti-ferromagnetic spin arrangements of BFO in the ${\cal R}$
  phase, calculated along one representative direction in the first
  BZ. Corresponding $\omega_{{\bf q}s}$ frequency pairs
  were identified by comparing the FM and G-AFM phonon eigenmodes.}
\label{fig2}
\end{figure}

\subsection{Spin-Phonon Couplings}
\label{subsec:spin-phonon}

We would like to identify a practical scheme to incorporate the main
effects of the spin fluctuations on the calculation of QH Helmholtz
free energies. To introduce our approach, let us begin by considering
the following general expression for the energy of the material, which
is the generalization of Eq.~(\ref{eq:eqh}) to the case of a compound
with localized magnetic moments whose interactions are well captured
by a Heisenberg Hamiltonian:
\begin{equation}
\begin{split}
E_{\rm harm}(\{u_{m}\},\{S_{i}\}) = & E^{0} + \frac{1}{2}
\sum_{mn}\Phi_{mn}^{0} u_{m} u_{n} \\ & + \frac{1}{2}\sum_{ij}
J_{ij}(\{u_{m}\}) S_{i} S_{j} \; ,
\end{split}
\label{eq:spin-phonon-energy}
\end{equation}
where the $S_{i}$ variables represent the magnetic moments associated
with specific atoms and the $J_{ij}$'s are the distortion-dependent
exchange interactions coupling them. (For brevity, in the following we
will talk about {\sl spins} instead of {\sl magnetic moments};
nonetheless, note that our arguments can be applied to cases involving
orbital magnetization.) To simplify the notation, we use complex
indexes -- $m$ and $n$ for the atomic displacements and $i$ and $j$
for the spins -- that include information about the cell, atom, and
Cartesian component defining the structural and magnetic
variables. Finally, we write the dependence of the exchange constants
on the atomic displacements as:
\begin{equation}
J_{ij}(\{u_{m}\}) = J_{ij}^{(0)} + \sum_{m} J_{ijm}^{(1)} u_{m} +
\frac{1}{2} \sum_{mn} J_{ijmn}^{(2)} u_{m} u_{n} \, ,
\end{equation}
where, for our purposes, it is sufficient to truncate the series at
the harmonic level. The $J_{ij}^{(0)}$ parameters describe the
magnetic interactions when the atoms remain at their equilibrium
positions; typically, these parameters will capture the bulk of the
exchange couplings. The $J_{ijm}^{(1)}$ coefficients describe the
forces that may appear on the atoms when we have certain spin
arrangements, and the $J_{ijmn}^{(2)}$ parameters capture the
dependence of the phonon spectrum on the spin configuration.

It is interesting to note that, while the energy in Eq.~(\ref{eq:eqh})
can be unambiguously described as a harmonic expansion around an
equilibrium state of the material, the interpretation of
Eq.~(\ref{eq:spin-phonon-energy}) is much more subtle. Indeed, because
we work with spin variables that have a fixed norm (nominally,
$|S_{i}| = 5\mu_{\rm B}$ in the case of the Fe$^{3+}$ cations in
BiFeO$_{3}$), the reference structure of our spin-phonon QH energy is
defined {\sl formally} as one in which the atomic spins are {\sl
  perfectly disordered} and the atoms are located at the corresponding
equilibrium positions. Such a structure cannot be easily considered in
a first-principles calculation; hence, we have to obtain the
parameters $E^{0}$ and $\Phi_{mn}^{0}$ that characterize it in an
indirect way. In essence, the fitting procedure would involve many
different spin configurations, and parameters $E^{0}$ and
$\Phi_{mn}^{0}$ would capture the part of the energy and
force-constant matrix that is {\sl independent} of the spin order.
Further, a thorough calculation of the $J_{ij}(\{u_{m}\})$ constants
would be a very challenging task. Indeed, a detailed modeling of the
spin-phonon couplings would require us to choose which spin pairs $i$
and $j$ are affected by which distortions pairs $m$ and $n$, a problem
that quickly grows in complexity even if we restrict ourselves to spin
interactions between first nearest neighbors.

In this work we did not attempt to pursue such a detailed description,
but adopted instead an approximate approach that provides the correct
results in particular important cases. To illustrate our scheme, let
us think of BFO's ${\cal R}$ phase and consider two specific spin
arrangements that are obviously relevant: (1) the G-AFM structure
(which is the ground state of the ${\cal R}$, ${\cal O}$, and ${\cal
  C}$ BFO phases mentioned above) and (2) a perfectly ferromagnetic
(FM) arrangement, which is the exact opposite case to G-AFM in the
sense that all the interactions between first nearest-neighboring
spins are {\sl reversed}. Let us also restrict ourselves to spin-spin
interactions between first nearest neighbors and, for the sake of
simplicity, let us assume that all first-nearest-neighboring spins are
coupled by the same $J$, so that we can drop the $i$ and $j$
indexes. (This is actually the case for the ${\cal R}$ phase of BFO,
and the generalization to other lower-symmetry cases is
straightforward.) Then, for a given spin arrangement $\gamma$ (where
$\gamma$ can be G-AFM or FM in this example), we can relax the atomic
structure of the material and construct the following energy
$E^{\gamma}_{\rm harm}$
\begin{equation}
E_{\rm harm}^{\gamma}(\{u_{m}\}) = E_{\rm eq}^{\gamma} + \frac{1}{2}
\sum_{mn}\Phi^{\gamma}_{mn} u_{m} u_{n} \; ,
\end{equation}
which is the analogous of Eq.~(\ref{eq:eqh}) above. Hence, we have
straightforward access to all the parameters in this expression from
first principles. Now, we want our general spin-phonon energy in
Eq.~(\ref{eq:spin-phonon-energy}) to reproduce $E^{\gamma}_{\rm harm}$
for the $\gamma$-orders of interest. If we are dealing with the G-AFM
and FM cases, it is trivial to check that this can be achieved by
making the following choices:
\begin{eqnarray}
&& E^{0} = \frac{1}{2}\left( E^{\rm FM}_{\rm eq} + E^{\rm G}_{\rm eq}
  \right) \, ,\\
&& \Phi_{mn}^{0} = \frac{1}{2}\left( \Phi_{mn}^{{\rm FM}} +
\Phi_{mn}^{{\rm G}} \right) \, ,\\
&& J^{(0)} = \frac{1}{6|S|^{2}}\left( E^{\rm FM}_{\rm eq} - E^{\rm
  G}_{\rm eq} \right) \, ,\\
&& J^{(1)}_{m} = 0 \, ,\\
&& J^{(2)}_{mn} = \frac{1}{6|S|^{2}}\left( \Phi_{mn}^{{\rm FM}} -
\Phi_{mn}^{{\rm G}} \right) \, .
\end{eqnarray}
While these choices may seem very natural, there are subtle
approximations and simplifications hiding behind them. For example,
the resulting model contains no explicit information about the atomic
rearrangements that may accompany a particular spin configuration;
nevertheless, the energies of the equilibrium structures are perfectly
well reproduced for the G-AFM and FM cases. Analogously, while the
phonons of the G-AFM and FM cases will be exactly reproduced by this
model, the spin-phonon interactions have been drastically simplified,
and we retain no information on how specific atomic motions affect
specific exchange constants. Hence, the resulting model should not be
viewed as an atomistic one; rather, it is closer to a phenomenological
approach in which we retain information about the effect of magnetic
order on the whole phonon spectrum.

Finally, we would like to use our spin-phonon energy to investigate
the properties of BFO at finite temperatures, especially in situations in
which the material is either paramagnetic (PM) or does not have a
fully developed AFM order. To do so, we will assume that, for the case
of fluctuating spins, the energetics of the atomic distortions is
approximately given by:
\begin{equation}
\begin{split}
 \tilde{E}_{\rm harm}(\{u_{m}\};x) & = E^{0} + 3x|S|^{2}J^{(0)} \\
& + \frac{1}{2} \sum_{mn} \left( \Phi_{mn}^{0} + 6 x|S|^{2}
 J^{(2)}_{mn}\right) u_{m}u_{n} \, ,
\end{split}
\label{eq:eqhprime4}
\end{equation}
where $x = \langle S_{i}S_{j} \rangle / |S|^{2}$ is the correlation
function between two neighboring spins, with $\langle ... \rangle$
indicating a thermal average. Note that in the limiting FM ($\langle
S_{i}S_{j} \rangle = |S|^{2}$) and G-AFM ($\langle S_{i}S_{j} \rangle
= -|S|^{2}$) cases, this equation reduces to the expected $E^{\rm
  FM}_{\rm harm}$ and $E^{\rm G}_{\rm harm}$ energies. Note also that
this model includes a spin-phonon contribution to the energy even in
the paramagnetic phase, as long as there are significant correlations
between neighboring spins. Indeed, for a non-zero value of $x$, the
phonon spectrum is given by the force-constant matrix
$\boldsymbol{\Xi}(x) \equiv \boldsymbol{\Phi}^{0} + 6 x|S|^{2}
\boldsymbol{J}^{(2)}$. 

In this work, we evaluate $x$ as a function of temperature by running
Monte Carlo simulations of the Heisenberg spin system described by the
$J^{(0)}_{ij}$ coupling constants, thus assuming frozen atomic
distortions. Then, since for a certain value of $x$
Eq.~(\ref{eq:eqhprime4}) is formally analogous to Eq.~(\ref{eq:eqh}),
we can apply the QH treatment described above to estimate the
Helmholtz free energy $\tilde{F}_{\rm harm}$ of the coupled
spin-phonon system.

Before concluding this section, let us discuss some approximate
expressions that can be obtained for $\tilde{F}_{\rm harm}$ and which
are illustrative of how our approach captures the effect of spin
fluctuations and of the peculiar nature of the paramagnetic state. We
have usually observed that the normal-mode frequencies
$\omega_{\boldsymbol{q}s}^{\gamma}$, obtained by diagonalizing the
dynamical matrix associated to $\boldsymbol{\Phi}^{\gamma}$, depend
significantly on the magnetic order. However, the corresponding
eigenvectors are largely independent from $\gamma$. As a result we
have the following approximate relations
\begin{eqnarray}
\tilde{\omega}_{{\bf q}s} \approx \omega_{{\bf q}s}^{\rm FM}
\sqrt{\left(\frac{\omega_{{\bf q}s}^{\rm G}}{\omega_{{\bf q}s}^{\rm
      FM}} \right)^{2} \left( \frac{1-x}{2} \right) + \left(
  \frac{1+x}{2} \right) } = \nonumber \\
\omega_{{\bf q}s}^{\rm G}
\sqrt{ \left( \frac{1-x}{2} \right) + \left(\frac{\omega_{{\bf
        q}s}^{\rm FM}}{\omega_{{\bf q}s}^{\rm G}} \right)^{2} \left(
  \frac{1+x}{2} \right)} \, ,
\label{eq:eqhprime6}
\end{eqnarray}
where $\{\tilde{\omega}_{\boldsymbol{q}s}\}$ are the frequencies
associated to $\boldsymbol{\Xi}(x)$. Using this result, we can write
the Helmholtz free energy in the high temperature limit as
\begin{eqnarray}
\tilde{F}_{\rm harm} (V,T \to \infty, \lbrace {\bf S} \rbrace) =
\left( \frac{F_{\rm harm}^{\rm FM} + F_{\rm harm}^{\rm G}}{2} \right)
+ \nonumber \\
\frac{3}{2} N k_{B} T \sum_{{\bf q}, s} {\rm ln} \left[
  \frac{\left(\omega_{{\bf q}s}^{\rm FM}\right)^{2} \left( 1 + x
    \right) + \left(\omega_{{\bf q}s}^{\rm G}\right)^{2} \left( 1 -x
    \right)}{2 \omega_{{\bf q}s}^{\rm FM} \cdot \omega_{{\bf
        q}s}^{\rm G} } \right]
\label{eq:eqhprime7}
\end{eqnarray}
where terms $F_{\rm harm}^{\rm FM}$ and $F_{\rm harm}^{\rm G}$ are
calculated in the classical limit through Eq.~(\ref{eq:classfharm})
and correspond to perfect FM and G-AFM spin arrangements. Note that in
the limiting cases $x = 1$ and $x=-1$, Eq.~(\ref{eq:eqhprime7})
consistently reduces to $F_{\rm harm}^{\rm FM}$ and $F_{\rm harm}^{\rm
  G}$. Interestingly, in the ideal paramagnetic case $x = 0$ we find
that, since all $\omega_{{\bf q}s}$ are positive, the inequality
$\tilde{F}_{\rm harm} > \frac{1}{2} \left( F_{\rm harm}^{\rm FM} +
F_{\rm harm}^{\rm G} \right)$ holds. This result sets a lower bound
for the error that we would be making if the spin disorder in ideal
paramagnetic phases was neglected. For example, in the case of the
${\cal R}$ phase of BFO, if we used a frozen G-AFM spin structure in
our QH calculations, the resulting free energy error would be of order
$\frac{1}{2} \left( F_{\rm harm}^{\rm FM} - F_{\rm harm}^{\rm G}
\right)$.

\subsection{Phonon Calculations}
\label{subsec:phonons}

In order to compute the QH free energy of a crystal, it is necessary
to know its full phonon spectrum over the whole BZ. There are
essentially two methods which can be used for the calculation of the
phonon frequencies: linear response theory and the direct
approach. The first method is generally exploited within the framework
of density functional perturbation theory
(DFPT);\cite{baroni87,baroni01,gonze89,gonze97} the main idea in DFPT
is that a linear order variation in the electron density upon
application of a perturbation to the crystal is the responsible for
the variation in the energy up to third order in the perturbation. If
the perturbation is a phonon wave with wave-vector ${\bf q}$, the
calculation of the density change to linear order can thus be used to
determine the dynamical matrix at wave-vector ${\bf q}$. This
procedure can be repeated at any wave-vector and without the need to
construct a supercell. In the direct approach, in contrast, the
force-constant matrix is directly calculated in real-space by
considering the proportionality between the atomic displacements and
forces when the former are sufficiently small (see
Sec.~\ref{subsec:Fqh}).\cite{kresse95,alfe01} In this case, large
supercells have to be constructed in order to guarantee that the
elements of the force-constant matrix have all fallen off to
negligible values at their boundaries, a condition that follows from
the use of periodic boundary conditions.\cite{alfe09} Once the
force-constant matrix is thus obtained, we can Fourier-transform it to
obtain the phonon spectrum at any $q$-point. In this work we chose to
perform phonon frequencies calculations with the direct
method. Nevertheless, convergence of the force-constant matrix
elements with respect to the size of the supercell in polar materials
may be slow due to the appearance of charge dipoles and macroscopic
electric fields in the limit of zero wave-vector; in the next section
we explain how we have efficiently dealt with this issue.

We performed a series of initial tests in the ${\cal R}$ phase to
determine the value of the various calculation parameters that
guarantee $F_{\rm harm}$ results converged to within $5$~meV/f.u. (As
it will be shown later, this accuracy threshold translates into
uncertainties of about 100~K in the predicted transition
temperatures.) The quantities with respect to which our QH free
energies need to be converged are the size of the supercell, the size
of the atomic displacements, and the numerical accuracy in the
calculation of the atomic forces and BZ sampling (see
Eq.~\ref{eq:fharm}).  We found the following settings to fulfill our
convergence requirements: $2 \times 2 \times 2$ supercells (i.e., 8
replications of the 10-atom unit cell of the ${\cal R}$ phase), atomic
displacements of $0.02$~\AA, and special
Monkhorst-Pack\cite{monkhorst76} grids of $12 \times 12 \times 12$
$q$-points, corresponding to the BZ of the ${\cal R}$-phase unit cell,
to compute the sums in Eq.~(\ref{eq:fharm}). Regarding the calculation
of the atomic forces with VASP, we found that the density of
$k$-points for BZ integrations had to be increased slightly with
respect to the value used in the energy calculations (e.g., from $8
\times 8 \times 8$ to $10 \times 10 \times 10$ for the BZ of the unit
cell of the ${\cal R}$ phase) and that computation of the non-local
parts of the pseudopotential contributions had to be performed in
reciprocal, rather than real, space. These technicalities were adopted
in all the phonon calculations, adapting in each crystal structure to
the appropriate $q$- and $k$-point densities.  The value of the phonon
frequencies and quasi-harmonic free energies were obtained with the
PHON code developed by Alf\`e.\cite{alfe09,phon} In using this code,
we exploited the translational invariance of the system to impose the
three acoustic branches to be exactly zero at the $\Gamma$ $q$-point,
and used central differences in the atomic forces (i.e., we considered
positive and negative atomic displacements). As an example of our
phonon frequency calculations, we show in Fig.~\ref{fig3} the full
phonon spectrum obtained for the ${\cal R}$ phase of BFO with a G-AFM
spin arrangement at zero pressure and when accounting for long-range
dipole-dipole interactions as described in the next section.

\subsection{Treatment of long-range Coulomb forces}
\label{subsec:lo-to}

As noted in the previous section, the displacement of atoms in an
insulator like BFO creates electric dipoles and long-range
dipole-dipole interactions; as a consequence, the interatomic force
constants $\Phi_{mn}$ decay typically with the third power of the
interatomic distance.  These long-range interactions play a critical
role in determining the spectrum of long-wavelength phonons. In the
direct approach, the phonon frequencies are exactly calculated at
wave-vectors ${\bf q}$ that are commensurable with the supercell;
thus, unaffordably large simulation boxes would in principle be needed
to accurately describe long-wavelength phonons.

Nevertheless, the long-range dipole-dipole interactions can be
modeled at the harmonic level from knowledge of the atomic Born
effective charge tensors and the dielectric tensor of the
material.\cite{gonze97,cochran62} Taking advantage of such a result,
Wang \emph{et al.} proposed a mixed-space approach in which accurate
force constants $\boldsymbol{\Phi}$ are calculated with the direct
approach in real space and long-range dipole-dipole interactions with
linear response theory in reciprocal space.\cite{wang10} Wang's
approach is based on the {\sl ad hoc} inclusion of a long-range
force-constant matrix of the form
\begin{equation}
\varphi_{l\kappa\alpha,l'\kappa'\alpha'} = \frac{4 \pi e^{2}}{N V}
\frac{\left( \sum_{\beta} q_{\beta} Z^{*}_{\kappa\beta,\alpha}\right)
  \left( \sum_{\beta} q_{\beta} Z^{*}_{\kappa\beta,\alpha'} \right)
}{\sum_{\beta\beta'} q_{\beta} \epsilon^{\infty}_{\beta\beta'} q_{\beta'}}
\label{eq:wang}
\end{equation}  
where $N$ is the number of primitive cells in the supercell and $V$
its volume; $e$ is the elemental charge,
$\boldsymbol{\epsilon}^{\infty}$ is the electronic dielectric tensor,
and $Z^{*}_{\kappa\beta,\alpha}$ is the Born effective charge
quantifying the polarization created along Cartesian direction
$\alpha$ when atom $\kappa$ moves along $\beta$. It can be shown that,
by Fourier-transforming the modified force-constant matrix
$\boldsymbol{\Omega} = \boldsymbol{\Phi} + \boldsymbol{\varphi}$, one
obtains the correct behavior near the $\Gamma$ point; further, for
${\bf q} \neq 0$ wave vectors one obtains a smooth interpolation that
recovers the exact results at the $q$-points commensurate with the
supercell employed for the calculation of
$\boldsymbol{\Phi}$.\cite{wang10}

\begin{figure}
\centerline{
\includegraphics[width=1.00\linewidth]{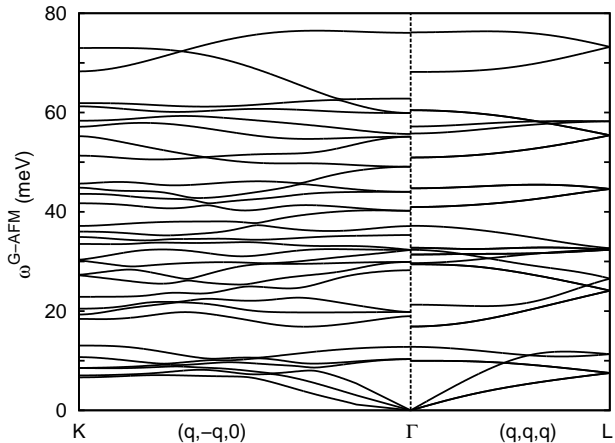}}
\caption{Phonon spectrum of the ${\cal R}$ phase of BFO with G-AFM
  spin order, calculated with the direct approach and considering
  long-range dipole-dipole interactions. The corresponding equilibrium
  volume per formula unit is 64.35~\AA$^{3}$.}
\label{fig3}
\end{figure}

In Table~I and Fig.~\ref{fig3}, we report the phonon frequencies that
we have obtained for the ${\cal R}$ phase of BFO using Wang's
mixed-space approach, and compare them to previous experimental and
theoretical data found in
Refs.~\onlinecite{hlinka11,borissenko13,hermet07}.  As it may be
appreciated there, the agreement between our $\Gamma$-phonon results
and the measurements is very good, indeed comparable to that achieved
with DFPT calculations performed by other authors.  (Actually, Wang's
method has already been applied with success to the study of the
phonon dispersion curves and the heat capacity of BFO.\cite{wang11})
After checking the numerical accuracy of Wang's technique, we
performed a test in which we assessed the $F_{\rm qh}$ differences
obtained by using the original and mixed-space versions of the direct
approach.  We found that the effect of considering long-range
dipole-dipole interactions in the QH energies was to vary $F_{\rm qh}$
in less than $5$~meV/f.u., which is our targeted accuracy
threshold. In view of the small size of these corrections, and for the
sake of computational affordability, we decided not to consider
$\boldsymbol{\varphi}$ terms in our subsequent calculations, for which
we just employed the original real-space version of the direct
approach.  In fact, as it has already been pointed out by
Alf\`e,\cite{alfe09} in the typical case an incorrect treatment of the
longitudinal optical modes near the $\Gamma$-point compromises only a
small region of the BZ, and the resulting errors in the free energy
are small and can in principle be neglected.

\begin{table*}
\begin{center}
\label{tab:lo-to-freq} 
\begin{tabular}{c c c c c c c c} 
\hline
\hline
$ $ & $ $ & $ $ & $ $ & $ $ & $ $ & $ $ & $ $ \\
$\quad {\rm TO~modes} \quad$ & $\quad {\rm This~work} \quad$ & $\quad {\rm Expt.} \quad$ & $\quad {\rm Calc.} \quad$ & $\quad {\rm LO~modes} \quad$ & $\quad {\rm This~work} \quad$ & $\quad {\rm Expt.} \quad$ & $\quad {\rm Calc.} \quad$ \\
$ $ & $ $ & $ $ & $ $ & $ $ & $ $ & $ $ & $ $ \\
\hline
$ $ & $ $ & $ $ & $ $ & $ $ & $ $ & $ $ & $ $ \\
$E({\rm TO1})$ & $78 $ & $74 $ & $102$ & $E({\rm LO1})$ & $85 $ & $81 $ & $104$ \\
$E({\rm TO2})$ & $136$ & $132$ & $152$ & $E({\rm LO2})$ & $161$ & $175$ & $175$ \\
$E({\rm TO3})$ & $238$ & $240$ & $237$ & $E({\rm LO3})$ & $242$ & $242$ & $237$ \\
$E({\rm TO4})$ & $252$ & $265$ & $263$ & $E({\rm LO4})$ & $258$ & $276$ & $264$ \\
$E({\rm TO5})$ & $265$ & $278$ & $274$ & $E({\rm LO5})$ & $323$ & $346$ & $332$ \\
$E({\rm TO6})$ & $330$ & $351$ & $335$ & $E({\rm LO6})$ & $352$ & $368$ & $377$ \\
$E({\rm TO7})$ & $361$ & $374$ & $378$ & $E({\rm LO7})$ & $393$ & $430$ & $386$ \\
$E({\rm TO8})$ & $412$ & $441$ & $409$ & $E({\rm LO8})$ & $445$ & $468$ & $436$ \\
$E({\rm TO9})$ & $488$ & $523$ & $509$ & $E({\rm LO9})$ & $483$ & $616$ & $547$ \\
$ $ & $ $ & $ $ & $ $ & $ $ & $ $ & $ $ & $ $ \\
$A_{1}({\rm TO1})$ & $151$ & $149$ & $167$ & $A_{1}({\rm LO1})$ & $172$ & $178$ & $180$ \\
$A_{1}({\rm TO2})$ & $219$ & $223$ & $266$ & $A_{1}({\rm LO2})$ & $240$ & $229$ & $277$ \\
$A_{1}({\rm TO3})$ & $285$ & $310$ & $318$ & $A_{1}({\rm LO3})$ & $461$ & $502$ & $428$ \\
$A_{1}({\rm TO4})$ & $506$ & $557$ & $517$ & $A_{1}({\rm LO4})$ & $550$ & $591$ & $535$ \\
$ $ & $ $ & $ $ & $ $ & $A_{2}({\rm LO1})$ & $101$ & $109$ & $109$ \\
$ $ & $ $ & $ $ & $ $ & $ $ & $ $ & $ $ & $ $ \\
\hline
\hline
\end{tabular} 
\end{center}
\caption{$\Gamma$-point phonon frequencies of the ${\cal R}$ phase of
  BFO with G-AFM spin order, calculated using the direct approach and
  considering long-range dipole-dipole interactions.  Experimental
  values are taken from Refs.~\onlinecite{hlinka11} and
  \onlinecite{borissenko13}, and previous LSDA-DFPT calculations from
  Ref.~\onlinecite{hermet07}. Frequencies are expressed in units of
  cm$^{-1}$.}
\end{table*}

\section{Results and Discussion}
\label{sec:results}

\subsection{Stability of the ${\cal R}$ and ${\cal O}$ phases at
  constant-volume and frozen-spin conditions}
\label{subsec:pnma}

\begin{figure}
\centerline{ \includegraphics[width=1.00\linewidth]{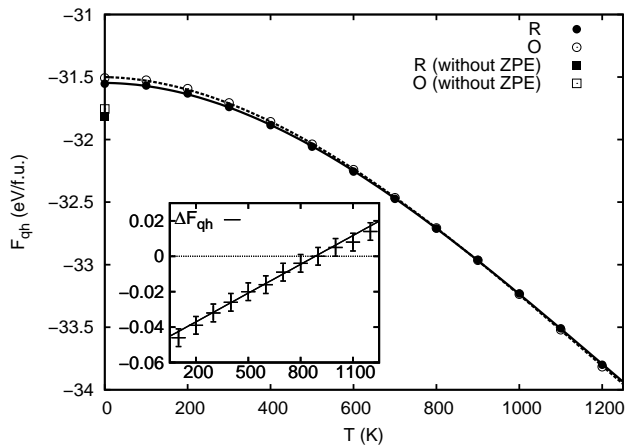}}
\caption{Quasi-harmonic free energies of the ${\cal R}$ and ${\cal O}$
  phases of BFO, calculated at $p \approx 0$~GPa (i.e., neglecting
  $T$-induced volume expansion effects) and considering a perfect
  G-AFM spin order in both structures. The size of the ZPE corrections
  is shown in the vertical axis. Inset: Plot of the quasi-harmonic
  free energy difference $\Delta F_{\rm qh} \equiv F_{\rm qh}({\cal
    R}) - F_{\rm qh}({\cal O})$ expressed as a function of
  temperature.}
\label{fig4}
\end{figure}

In this section, we present our QH results for the ${\cal R}$ and
${\cal O}$ phases of BFO. For the sake of clarity, we first discuss
the results obtained when spin-disorder and volume-expansion effects
are neglected.

In Fig.~\ref{fig4} we plot the $F_{\rm qh}$ energy of the ${\cal R}$
and ${\cal O}$ phases calculated at $p \approx 0$~GPa as a function of
temperature. Volumes were kept fixed at their equilibrium values
$V_{0}$ obtained at $T = 0$~K, which are equal to
64.61~\AA$^{3}$/f.u. and 61.99~\AA$^{3}$/f.u., respectively. We
considered the perfect G-AFM spin order to be frozen in both
structures. We computed $F_{\rm qh}$ over sets of fourteen temperature
points taken at intervals of 100~K and fitted them to third-order
polynomial curves. ZPE corrections (see Eq.~\ref{eq:zpe}) were
included in the fits and are equal to 0.263(5)~eV/f.u. and
0.248(5)~eV/f.u., respectively, for the ${\cal R}$ and ${\cal O}$
phases. (An estimate of the error is given within parentheses.) We
find that at $T = 0$~K the ${\cal R}$ phase is energetically more
favorable than the ${\cal O}$ phase by 0.046(5)~eV/f.u.  As the
temperature is raised, however, the Helmholtz free energy of the
${\cal O}$ phase becomes lower than that of the ${\cal R}$ phase due
to the increasingly more favorable $F_{\rm harm}$ contributions. For
instance, at $T = 300$~K, $F_{\rm harm}$ amounts to
0.048(5)~eV/f.u. for the ${\cal O}$ phase and 0.077(5)~eV/f.u. for the
${\cal R}$ phase, whereas at $T = 1000$~K the obtained values are
$-$1.481(5)~eV/f.u. and $-$1.414(5)~eV/f.u.,
respectively. Consequently, a first-order phase transition of the
${\cal R} \to {\cal O}$ type is predicted to occur at
$T_{t}$~=~900(100)~K. We show this in the inset of Fig.~\ref{fig4},
where the energy difference $\Delta F_{\rm qh} \equiv F_{\rm qh}({\cal
  R}) - F_{\rm qh}({\cal O})$ is represented as a function of
temperature; since quasi-equilibrium conditions are assumed, the
corresponding transition temperature coincides with the point at which
$\Delta F_{\rm qh} = 0$.  We notice that this estimation of $T_{t}$ is
reasonably close to the experimental value of
1100~K.\cite{arnold09,catalan09}

\begin{figure}
\centerline
        {\includegraphics[width=1.0\linewidth]{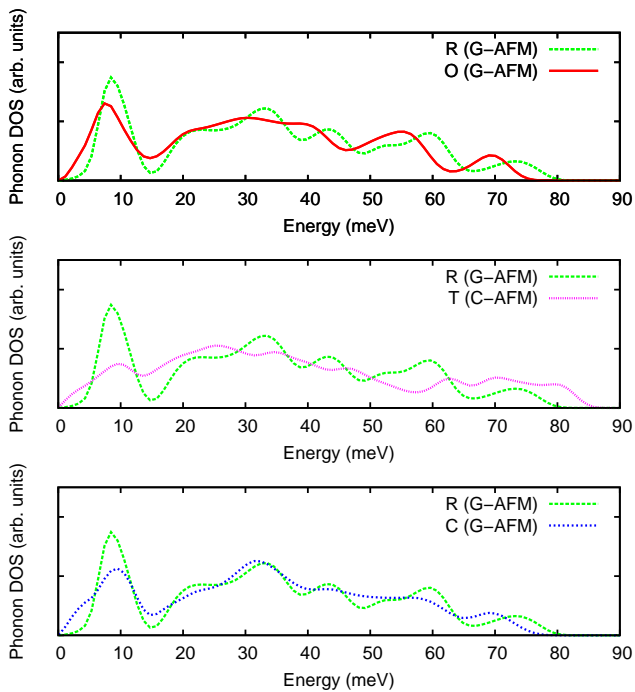}}%
\caption{(Color online) Phonon density of states of various BFO phases obtained at 
  $p = 0$~GPa. The corresponding equilibrium volumes are
  64.61~\AA$^{3}$/f.u., 61.99~\AA$^{3}$/f.u., 71.12~\AA$^{3}$/f.u.,
  and 64.17~\AA$^{3}$/f.u. for the ${\cal R}$, ${\cal O}$, ${\cal T}$,
  and ${\cal C}$ phases, respectively. }
\label{fig5}
\end{figure}

Let us now discuss the origin of the obtained solid-solid
transformation in terms of the phonon eigenmodes and frequencies of
each phase. In Fig.~\ref{fig5} we plot the phonon density of states
(pDOS) calculated for the ${\cal R}$ and ${\cal O}$ phases at their
equilibrium volumes. We find that the value of the geometric frequency
$\bar{\omega}$ (see Eq.~\ref{eq:omegabar}) is 27.16~meV in the ${\cal
  O}$ phase and 28.58~meV in the ${\cal R}$ structure (expressed in
units of $\hbar$). Therefore, as it was already expected from the
results shown in Fig.~\ref{fig4}, the ${\cal O}$ phase of BFO is, in
average, vibrationally softer than the ${\cal R}$ phase. In
particular, the pDOS of the ${\cal O}$ phase accumulates a larger
number of phonon modes within the low-energy region of the spectrum,
and extends over a smaller range of frequencies.

\begin{figure}
\centerline{
\includegraphics[width=1.00\linewidth]{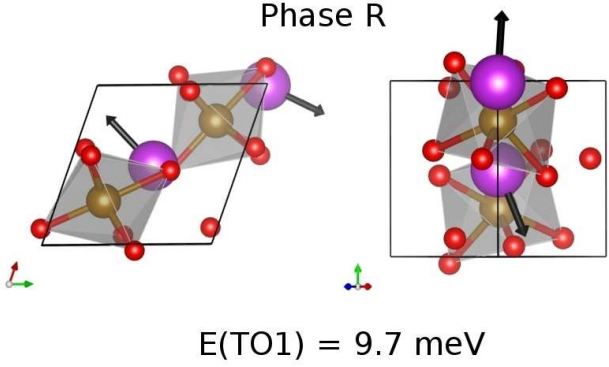}}
\vspace{-0.50cm}
\caption{(Color online) Sketch of the first optical transverse $\Gamma$-point 
  phonon mode obtained in the ${\cal R}$ phase of BFO at equilibrium. Bi
  displacements are represented with black arrows, and Bi, Fe, and O
  atoms with purple, brown and red spheres, respectively.}
\label{fig6}
\end{figure}

\begin{figure}
\centerline{
\includegraphics[width=1.00\linewidth]{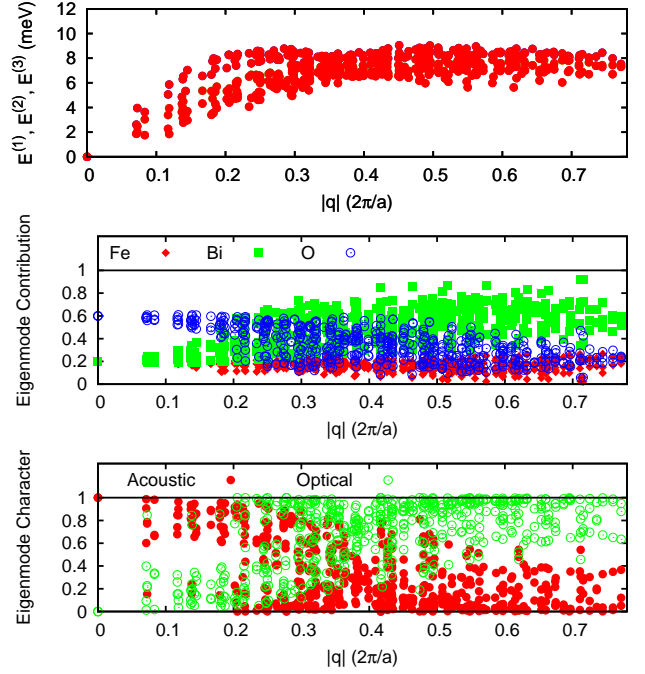}}
\caption{(Color online) Analysis of the three lowest-energy phonon eigenmodes
  obtained at the $q$-points used for the sampling of the BZ of BFO's
  ${\cal R}$ phase.  We represent their eigenenergies as a function of
  wave-vector module in the top panel, the contribution of each atomic
  species to the mode eigenvectors in the middle panel, and a
  quantification of their acoustic and optical characters in the
  bottom panel. For this quantification, we took advantage of the
  normalization and orthogonality relations satisfied by the
  eigenvectors of the dynamical matrix calculated at $\Gamma$ and
  ${\bf q} \neq 0$ points.}
\label{fig7}
\end{figure}

We restrict our following analysis to the low-energy phonons (i.e.,
$\omega_{{\bf q}s} \le \bar{\omega}$), which provide the dominant
contributions to $F_{\rm harm}$.  In the ${\cal R}$ phase, we observe
a sharp pDOS peak centered at $\hbar \omega \sim 10$~meV followed by a
deep valley. By inspecting the spectrum of phonon eigenmodes obtained
at $\Gamma$ and the full phonon bands displayed in Fig.~\ref{fig3}, we
identify that pDOS maximum with the first optical transverse mode TO1
(see Fig.~\ref{fig6}). This phonon mode involves opposed displacements
of neighboring Bi atoms within the plane perpendicular to the
pseudo-cubic direction $[111]$, and is polar in the $[10\bar{1}]$
direction.\cite{aclaration-polarity} Figure~\ref{fig7} gives
additional information on the three lowest-lying phonons of the ${\cal
  R}$ phase across the BZ. There we can see that the softest phonons
are acoustic in character and correspond to $q$-points in the
neighborhood of $\Gamma$. As we move away from $\Gamma$, the
lowest-lying phonon modes change of character and can be represented
by the optical distortion shown in Fig.~\ref{fig6}.

\begin{figure}
\centerline{
\includegraphics[width=1.00\linewidth]{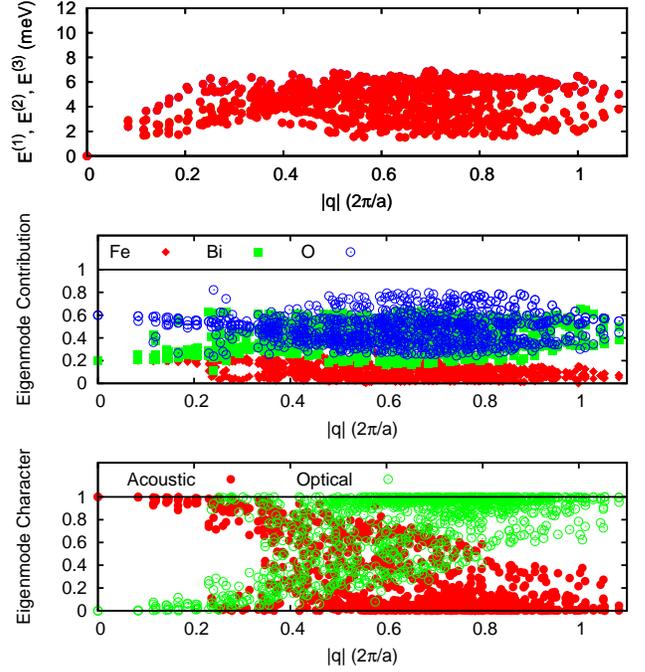}}
\caption{(Color online) Same as Fig.~\protect\ref{fig7}, but for BFO's ${\cal O}$
  phase.}
\label{fig8}
\end{figure}

The situation for the ${\cal O}$ phase is rather different. As it can
be appreciated from Fig.~\ref{fig5}, the number of phonons in the very
low-frequency region is much greater than in the ${\cal R}$ phase.
Small frequency values are in general related to phonon modes of
strong acoustic character, which are the responsible for the elastic
response of materials: the softer a crystal is, the smaller the slopes
of its acoustic bands around the $\Gamma$ $q$-point, and the larger
the number of low-energy phonons that result. By applying this
reasoning to the present case and considering our pDOS results, one
would arrive at the conclusion that BFO in the ${\cal O}$ phase should
be elastically softer than in the ${\cal R}$ phase. However, this is
not the case: we computed the equilibrium bulk modulus of BFO (i.e.,
$B \equiv -V \frac{\partial p}{\partial V}$) at $T = 0$~K describing
the response of the material to uniform deformations and found,
respectively, $99(2)$~GPa and $158(2)$~GPa for the ${\cal R}$ and
${\cal O}$ structures. Interestingly, this apparent contradiction is
quickly resolved by inspecting the behavior of the (three)
lowest-lying phonons calculated at each BZ $q$-point (see
Fig.~\ref{fig8}). As clearly observed in Fig.~\ref{fig8}, the ${\cal
  O}$ phase of BFO presents very low-lying optical bands with phonon
frequencies that can be below 2~meV. The corresponding eigenmodes are
dominated by the stretching of Bi--O bonds, with the Fe ions having a
very minor contribution (see middle panel in Fig.~\ref{fig8}).  In
fact, these soft optical phonons, with ${\bf q}$ vectors far away from
$\Gamma$, are the ones responsible for the stabilization of BFO's
${\cal O}$ phase at high temperatures.

\subsection{Effect of spin disorder on the ${\cal R} \to {\cal O}$ transition}
\label{subsec:m-effects}

\begin{figure}
\centerline{ \includegraphics[width=1.00\linewidth]{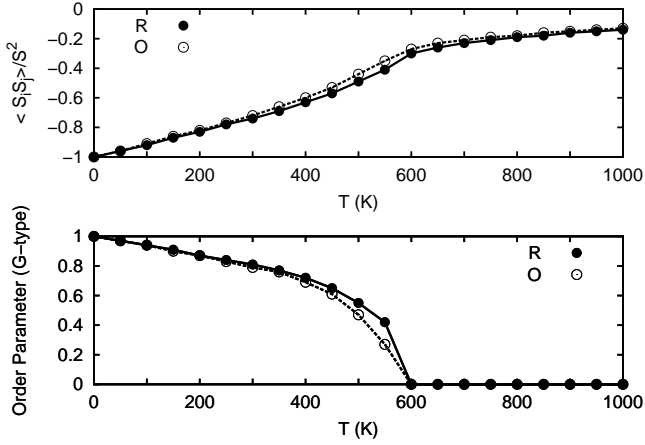}}
\caption{Monte Carlo results obtained for a simple Heisenberg model
  reproducing the spin magnetic order in the ${\cal R}$ and ${\cal O}$
  phases of BFO. Top: Average value of the normalized spin product
  ${\bf S}_{i}\cdot{\bf S}_{j}$ (with $S \equiv 5/2 \mu_{\rm B}$) as a
  function of temperature.  Bottom: Calculated order parameter $S_{\rm
    G}$ (see text) as a function of temperature.}
\label{fig9}
\end{figure}

\begin{figure}
\centerline{
\includegraphics[width=1.00\linewidth]{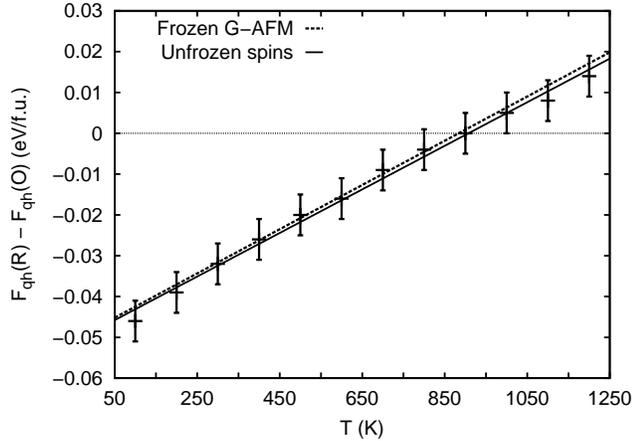}}
\caption{Quasi-harmonic free energy difference between the ${\cal R}$
  and ${\cal O}$ phases of BFO. We show the results obtained in two
  different situations, i.e., frozen G-AFM and $T$-dependent spin
  orders (solid symbols and error bars correspond to the last case).
  Lines are linear fits to the free energy results.}
\label{fig10}
\end{figure}

In order to assess the effect of spin fluctuations on the predicted
${\cal R} \to {\cal O}$ phase transition, we put in practice the ideas
explained in Sec.~\ref{subsec:spin-phonon}. As described there, our
practical approach to capture the effects of spin disorder requires
the calculation of the QH energies for the G-AFM ($F_{\rm qh}^{\rm
  G}$; this is the case already considered in the previous section)
and FM ($F_{\rm qh}^{\rm FM}$) spin arrangements, from which we derive
the parameters describing (1) the spin-independent part of the energy
($E^{0}$ and $\boldsymbol{\Phi}^{0}$), (2) the Heisenberg spin
Hamiltonian for zero atomic distortions ($J^{0}$), and (3) the effects
of the spin arrangement on the phonon spectrum
($\boldsymbol{J}^{(2)}$). Our DFT calculations render $J^{(0)}$ values
of 34.67~meV and 32.67~meV, respectively, for the ${\cal R}$ and
${\cal O}$ phases, indicating a similar and strong tendency towards
the G-AFM order. Further, Fig.~\ref{fig2} shows illustrative
results of the shifts in phonon frequencies, for the ${\cal R}$ phase
of BFO, that occur as a function of the spin structure; these are the
effects captured by the $\boldsymbol{J}^{(2)}$ terms.

Figure~\ref{fig9} reports the results of a series of Monte Carlo (MC)
simulations performed with the Heisenberg model defined by the
$J^{(0)}$ coupling. We used a periodically-repeated simulation box of
$20 \times 20 \times 20$ spins, and computed the thermal averages from
runs of 50000 MC sweeps. The aim of these simulations was to determine
the value of the spin average $\langle {\bf S}_{i} \cdot {\bf S}_{j}
\rangle$ that has to be used in
Eqs.~(\ref{eq:eqhprime4})-(\ref{eq:eqhprime6}) and which depends on
$T$. Note that here we have abandoned the compact notation of
Section~\ref{subsec:spin-phonon}, and $S_{i\alpha}$ denotes the
$\alpha$ Cartesian component of the spin at cell $i$. Additionally,
these simulations allow us to monitor the occurrence of magnetic
transitions through the computation of the G-AFM order parameter
$S^{\rm G} = \frac{1}{N} \sum_{i} (-1)^{n_{ix}+n_{iy}+n_{iz}}
S_{iz}$. Here, $n_{ix}$, $n_{iy}$, and $n_{iz}$ are the three integers
locating the $i$-th lattice cell, and $N$ is the total number of spins
in the simulation box; further, for the calculation of $S^{\rm G}$, we
need to consider only the $z$ component of the spins because of a
small symmetry-breaking magnetic anisotropy that was included in our
Hamiltonian to facilitate the analysis (see Supplemental Material of
Ref.~\onlinecite{escorihuela12}). Our results show that in the ${\cal
  R}$ phase the magnetic phase transition occurs at $T \sim 600$~K, a
temperature that is rather close to the experimental value $T_{N} \sim
650$~K.\cite{kiselev63,smolenskii61,catalan09} The results for the
${\cal O}$ phase are very similar.

Now, let us assess the consequences of considering these effects on
the QH free energies of the ${\cal R}$ and ${\cal O}$ phases (see
Eqs.~\ref{eq:eqhprime4}-\ref{eq:eqhprime6}).  Figure~\ref{fig10}
reports the free energy difference between the ${\cal R}$ and ${\cal
  O}$ phases, as obtained by considering ($\Delta \tilde{F}_{\rm qh}$)
or neglecting ($\Delta F_{\rm qh}$) the effect of the spin
fluctuations.  As one can appreciate, the two curves are almost
identical and provide the same transition temperature. At $T = 300$~K,
for instance, both $\Delta F^{\rm G}_{\rm qh}$ and $\Delta
\tilde{F}_{\rm qh}$ are about $-$0.032(5)~eV/f.u., and at $T = 1000$~K
we get approximately 0.006(5)~eV/f.u.; that is, the differences fall
within the accuracy threshold set in our free energy calculations.

However, the consequences of considering $T$-dependent spin
arrangements in the calculation of the QH free energy of an individual
phase are actually quite sizable. Indeed, the error function defined
as $\delta \tilde{F}_{\rm qh} \equiv \tilde{F}_{\rm qh} - F^{\rm
  G}_{\rm qh}$ may amount to several tenths of eV at high
temperatures.  For instance, for the ${\cal R}$ phase, $\delta
\tilde{F}_{\rm qh}$ is 0.068(5)~eV/f.u. at $T = 300$~K and
0.102(5)~eV/f.u. at $T = 1000$~K.  Hence, the reason behind the
numerical equivalence between functions $\Delta F^{\rm G}_{\rm qh}$
and $\Delta \tilde{F}_{\rm qh}$ is that $\delta \tilde{F}_{\rm qh}$
errors are essentially the same in both ${\cal R}$ and ${\cal O}$
structures and thus they cancel.  Consequently, it is possible to
obtain reasonable $T_{t}$ predictions in BFO even if one neglects the
strong dependence of spin magnetic order on temperature.

In view of this conclusion, and for the sake of computational
affordability, we will disregard spin-disorder effects for the rest of
phases considered in this work. Also, we note that the inequality
$\tilde{F}_{\rm harm} > \frac{1}{2} \left( F_{\rm harm}^{\rm FM} +
F_{\rm harm}^{\rm G} \right)$, mentioned in
Sec.~\ref{subsec:spin-phonon}, is fulfilled for both the ${\cal R}$
and ${\cal O}$ structures at all temperatures, even when $x = \langle
{\bf S}_{i} \cdot {\bf S}_{j} \rangle / |S|^{2} \neq 0$ and $\hbar
\bar{\omega} \approx k_{B} T$.  Plausibly then, the lower bound set
there for the QH errors caused by neglecting the spin disorder can be
tentatively generalized to any value of $x$.

\begin{figure}
\centerline{
\includegraphics[width=1.00\linewidth]{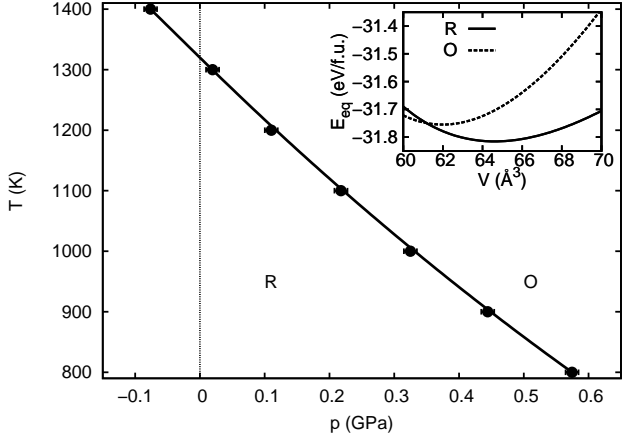}}
\caption{Calculated two-phase boundary delimiting the ${\cal R}$ and
  ${\cal O}$ regions in the bulk phase diagram of BFO at elevated
  temperatures. The solid line is a guide to the eyes and the symbols
  represent explicitly calculated points. Inset: $E_{\rm eq}$ curves
  obtained for the ${\cal R}$ and ${\cal O}$ phases as a function of
  volume without considering ZPE corrections.}
\label{fig11}
\end{figure}

\begin{figure}
\centerline{
\includegraphics[width=1.00\linewidth]{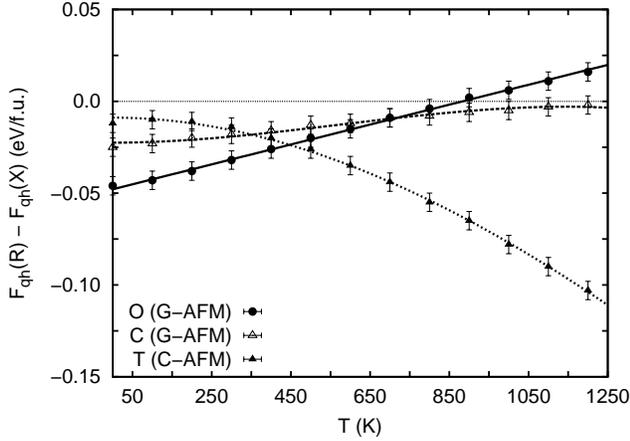}}
\caption{Quasi-harmonic free energy differences among the ${\cal R}$
  phase and the rest of crystal structures analyzed in this
  work. Perfect G-AFM spin order and constrained equilibrium volumes
  were considered. Lines are guides to the eyes and the symbols
  represent explicitly calculated points.}
\label{fig12}
\end{figure}

\subsection{Effect of volume expansion on the ${\cal R} \to {\cal O}$
  transition}
\label{subsec:volume-effects}

To address the effect of volume expansion on $T_{t}$, we performed
additional energy, phonon, and $F_{\rm qh}$ calculations over a grid
of five volumes spanning the interval $0.95 \le V/V_{0} \le 1.10$ for
both ${\cal R}$ and ${\cal O}$ phases. At each volume, first we
computed the value of $F_{\rm qh}$ at a series of temperatures in the
range between 0~K and 1600~K, taken at $100$~K intervals. Then, at
each $T$ we fitted the corresponding $F_{\rm qh} (V, T)$ points to
third-order Birch-Murnaghan equations\cite{birch78,cazorla13} and
performed Maxwell double-tangent constructions over the resulting
${\cal R}$ and ${\cal O}$ curves to determine $p_{t} (T)$ (i.e., the
pressure at which the first-order ${\cal R} \to {\cal O}$ transition
occurs at a given $T$). By repeating this process several times we
were able to draw the ${\cal R}$--${\cal O}$ phase boundary,
$p_{t}(T)$, in the interval $-$0.1~GPa~$\le p \le
0.6$~GPa. Figure~\ref{fig11} reports these results. As one can
appreciate there, the calculated transition temperature at equilibrium
now is 1300(100)~K. Volume expansion effects, therefore, shift upwards
by 400~K our previous tentative $T_{t}$ estimation. Also, we find that
the volume of the crystal varies from 66.51~\AA$^{3}$/f.u. to
62.34~\AA$^{3}$/f.u. during the course of the ${\cal R} \to {\cal O}$
transformation.  These values can be compared to recent experimental
data obtained by Arnold \emph{et al.}\cite{arnold09} which are $T_{\rm
  C} \approx 1100$~K, $V({\cal R}) = 64.15$~\AA$^{3}$/f.u. and
$V({\cal O}) = 63.10$~\AA$^{3}$/f.u. In general, our agreement with
respect to Arnold's measurements can be regarded as reasonably good,
although our QH calculations overestimate the transition temperature
and volume reduction $\Delta V = V({\cal O})-V({\cal R})$ observed in
experiments.

Moreover, in the vicinity of the transition state [0~GPa,
  1300(100)~K], we assumed the slope of the phase boundary to be
constant and numerically computed $dT/dp \approx -$1100~K/GPa. By
introducing this value and $\Delta V = -$4.17~\AA$^{3}$/f.u. in the
Clausius-Clapeyron equation, we found the latent heat of the
ferroelectric phase transformation to be about
0.71~Kcal/mol. Unfortunately, we do not know of any experimental data
to compare this result with.  Interestingly, if we assume the slope of
the ${\cal R}$--${\cal O}$ phase boundary to be constant regardless of
the $p-T$ conditions, the extrapolated zero-temperature ${\cal R} \to
{\cal O}$ transition turns out to be $p_{t} (0) \sim 1.2$~GPa. This
result differs greatly from the $p_{t} (0~K)$ value obtained when
straightforwardly considering static $E_{\rm eq}$ curves (see inset of
Fig.~\ref{fig11}) and enthalpies (i.e., $H_{\rm eq} = E_{\rm eq} +
p_{\rm eq} V$), which is 4.8~GPa. This disagreement may indicate that
assuming global linear behavior in $p_{t} (T)$ is unrealistic and/or
that ZPE corrections in BFO are very important. We will comment again
on this point in Sec.~\ref{subsec:p-induced}, when analyzing in detail
the role of ZPE corrections in prediction of $p$-induced phase
transformations at $T = 0$~K.

\subsection{The complex ${\cal C}$ phase}
\label{subsec:complex}

\begin{figure}
\centerline{\includegraphics[width=1.00\linewidth]{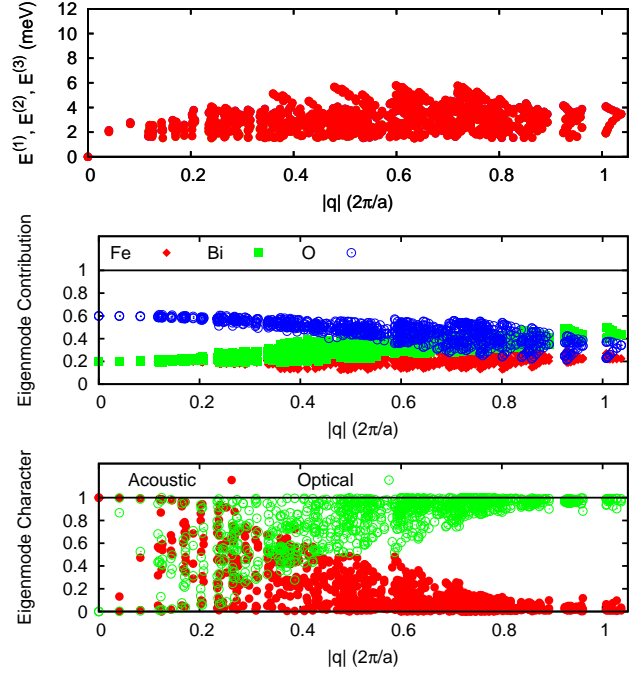}}
\caption{(Color online) Same as Fig.~\protect\ref{fig7}, but for the complex ${\cal
    C}$ phase considered in this work.}
\label{fig13}
\end{figure}

Novel \emph{nanoscale-twinned} structures, denoted here as {\sl
  complex} or ${\cal C}$ phases, have been recently suggested to
stabilize in bulk BFO under conditions of high-$T$ or high-$p$, and
upon appropriate chemical substitutions.\cite{prosandeev12} From a
structural point of view, these ${\cal C}$ phases can be thought of a
{\sl bridge} appearing whenever we have ${\cal R}$ and ${\cal O}$
regions in the phase diagram of BFO. Thus, the energies of the
nanoscale-twinned structures lie very close to that of the ground
state.  Interestingly, these ${\cal C}$ phases have been linked also
to the structure of domain walls whose energy is essentially
determined by antiferrodistortive modes involving the rotation of
O$_{6}$ octahedra.\cite{dieguez13} These intriguing features
motivated us to study the thermodynamic stability of this new type of
phases with the QH approach.  We note that, in the original paper by
Prosandeev \emph{et al.},\cite{prosandeev12} several phases are
proposed as members of the ${\cal C}$ family. For reasons of
computational affordability, we restrict our analysis here to one
particular structure (with $Pca2_{1}$ space group and $Z = 8$) that
has been introduced above and is depicted in Fig.~\ref{fig1}(d).

In Fig.~\ref{fig12} we plot the QH free energy of our ${\cal C}$ phase
expressed as a function of temperature, taking the result for the
${\cal R}$ structure as the zero of energy. As one may observe there,
at low $T$ the $\Delta F_{\rm qh}$ difference is negative and quite
small in absolute value.  At $T = 0$~K, for instance, this quantity
amounts to $-$0.025~(5)~eV/f.u., and roughly lies between the values
corresponding to the ${\cal T}$ and ${\cal O}$ phases. As $T$ is
raised, however, $\Delta F_{\rm qh}$ increases steadily with an
approximate slope of $2 \times 10^{-5}$~eV/K, and at $T \approx
1000$~K it becomes positive within our numerical uncertainties. This
change of sign marks the occurrence of a potential ${\cal R} \to {\cal
  C}$ transformation. However, such a transition would be prevented by
the onset of the ${\cal O}$ phase, which becomes the equilibrium state
at a lower temperature. Note that, according to our results, the
prevalence of the ${\cal O}$ phase occurs in spite of the fact that,
at 0~K, this phase is energetically less favorable than the ${\cal
  C}$ state by 0.021~eV/f.u.

The free energy competition between the ${\cal O}$ and ${\cal C}$
phases is very strong, as can be deduced from the pDOS plots enclosed
in Fig.~\ref{fig5}. In particular, the ${\cal C}$ phase shares common
pDOS features with both the ${\cal R}$ and ${\cal O}$ structures,
which is hardly surprising given that its atomic arrangement can be
viewed as a mixture between the ${\cal R}$ and ${\cal O}$
solutions. For instance, in the $\omega \to 0$ limit, the ${\cal C}$
and ${\cal O}$ distributions are practically identical, and the range
of phonon frequencies over which they expand is very
similar. Moreover, the number of low-lying optical phonon modes found
in the ${\cal C}$ phase is, as we calculated for the ${\cal O}$
structure, very high (although we note that in the ${\cal C}$ case the
contribution of the Fe anions to the eigenmodes is not negligible, see
Fig.~\ref{fig13}). Then, for intermediate frequencies the ${\cal C}$
pDOS presents a series of modulations which are more characteristic of
the ${\cal R}$ phase. Also, the energy of the first ${\cal C}$ pDOS
peak is closer to that of the ${\cal R}$ phase, and from an elastic
point of view both ${\cal C}$ and ${\cal R}$ phases are very similar
(that is, the bulk modulus of the two structures are coincident within
our numerical uncertainties).  A quantitative testimony of these pDOS
similarities is given by the geometric frequencies $\bar{\omega}$
calculated in the ${\cal O}$, ${\cal C}$, and ${\cal R}$ phases, which
are 27.16~meV, 28.00~meV, and 28.58~meV, respectively. Furthermore,
ZPE corrections in the ${\cal C}$ phase amount to 0.254~eV/f.u., a
value that roughly coincides with the arithmetic average obtained for
the corresponding ${\cal O}$ and ${\cal R}$ results.  In conclusion,
we can state that BFO in the ${\cal C}$ phase is in average
vibrationally softer than in the ${\cal R}$ phase, but more rigid than
in the ${\cal O}$ phase.
  
It is worth noticing that, although we do not predict here a
temperature-induced phase transition of the ${\cal R} \to {\cal C}$
type, this can not be discarded to occur in practice given that the
calculated $\Delta F_{\rm qh}$ differences among the ${\cal R}$,
${\cal O}$ and ${\cal C}$ structures are very small. Note that small
variations in the computed free energies -- as for instance due to the
use of a different exchange-correlation functional in our DFT
calculations, related to our QH approximation, etc. -- could very well
change this delicate balance of relative stability (see discussion
in Sec.~\ref{subsec:lda}). Further, the ${\cal C}$ phase considered
here is only one among the many nanoscale-twinned structures that have
been predicted to exist,\cite{prosandeev12} and it is reasonable to
speculate that some of them might indeed be predicted to be the
equilibrium solutions by the DFT scheme employed here. At any rate,
our results do suggest that these ${\cal C}$ structures are, at the
very least, very close to becoming stable in the regions of the phase
diagram in which ${\cal R} \to {\cal O}$ transitions occur. Moreover,
they are obvious candidates to {\sl mediate} (i.e., to appear in the
path of) the ${\cal R} \to {\cal O}$ transformation. Hence, our
results are clearly compatible with the possibility that ${\cal C}$
phases can be accessed experimentally, as robust meta-stable states,
depending on kinetic factors.

\subsection{The super-tetragonal ${\cal T}$ phase}
\label{subsec:cc}

\begin{figure}
\centerline{\includegraphics[width=1.00\linewidth]{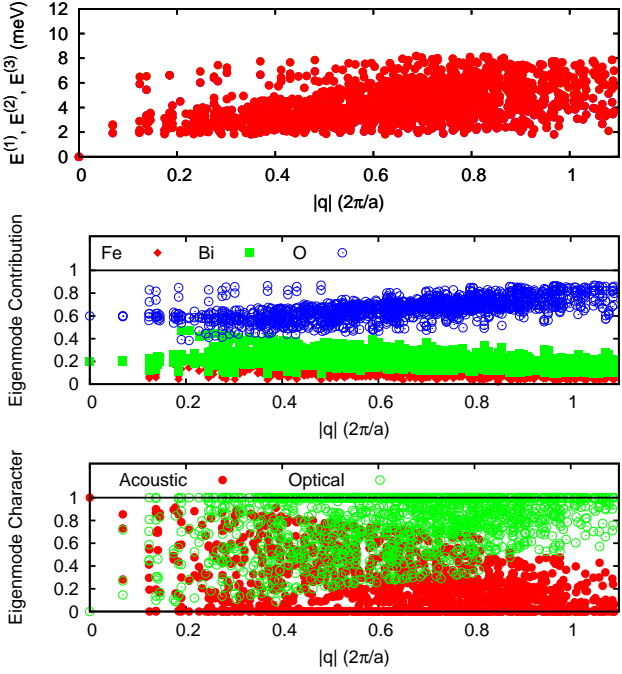}}
\caption{(Color online) Same as Fig.~\protect\ref{fig7}, but for the super-tetragonal
  ${\cal T}$ phase considered in this work.}
\label{fig14}
\end{figure}

Under zero $p-T$ conditions, the energy of the ${\cal T}$ structure
depicted in Fig.~\ref{fig1}(d) differs from that of the ${\cal R}$
phase by only few hundredths of eV per formula unit.\cite{dieguez11}
This ${\cal T}$ phase possesses a giant $c/a$ ratio, a large electric
polarization with a small in-plane component, and anti-ferromagnetic
spin order of type C (C-AFM); hence, in principle, this phase would be
potentially relevant for technological applications. Nevertheless, the
${\cal T}$ phase has never been observed in bulk samples of BFO
(although it is stabilized in thin films under high compressive and
tensile epitaxial constraint\cite{bea09,zeches09}). Aiming at
understanding the causes behind the frustrated stabilization of a
bulk-like ${\cal T}$ phase in BFO, we studied it with the QH approach.

In Fig.~\ref{fig12} we plot the QH free energy of the ${\cal T}$ phase
taken with respect to that of the ${\cal R}$ structure and expressed
as a function of temperature.  The ${\cal T}$ phase is assumed to
present frozen C-AFM spin order, and a frozen G-AFM arrangement is
considered for the ${\cal R}$ phase. As one may observe there, the
free energy difference $\Delta F_{\rm qh} (T)$ is negative and very
small at low temperatures (e.g., $\Delta F_{\rm
  qh}$(0~K)~=~$-$0.012~(5)~eV/f.u.) but progressively increases in
absolute value as $T$ is raised (e.g., $\Delta F_{\rm
  qh}$(1000~K)~=~$-$0.078~(5)~eV/f.u.). This result implies that
vibrational thermal excitations energetically destabilize the ${\cal
  T}$ phase as compared to the ${\cal R}$ and ${\cal O}$ structures,
in agreement with observations.

This conclusion may not seem so obvious from inspection of the pDOS
results enclosed in Fig.~\ref{fig5}. As we can see there, at
frequencies below 5~meV, the ${\cal T}$ phase presents a larger phonon
density than the ${\cal R}$ phase, which would in principle suggest
that the ${\cal T}$ structure is vibrationally softer. However, the
lowest-lying pDOS peak in the ${\cal R}$ phase is much higher than in
the ${\cal T}$ structure, and this feature turns out to be
dominant. In particular, the calculated geometric frequency
$\bar{\omega}$ amounts to 33.07~meV in the ${\cal T}$ phase and to
28.58~meV in the ground state.  Interestingly, ZPE corrections (see
Eq.~\ref{eq:zpe}) in both ${\cal R}$ and ${\cal T}$ phases are
practically identical ($\sim$~0.26~eV/f.u.).

The relatively high number of phonon modes that the ${\cal T}$ phase
presents at very low frequencies is reminiscent of the results
discussed above for the ${\cal O}$ structure. Indeed, as can be seen
in Fig.~\ref{fig14}, in the ${\cal T}$ phase we also find low-lying
phonons of very low energy throughout the BZ. Additionally, the ${\cal
  T}$ phase also presents a relatively small bulk modulus and is
elastically softer than the ${\cal R}$ structure: we obtained
73(2)~GPa in this case, while we calculated 99(2)~GPa for the ${\cal
  R}$ phase. These bulk modulus results are consistent with what one
would generally expect from inspection of the pDOS plots enclosed in
Fig.~\ref{fig5}; in this sense, the ${\cal T}$ structure behaves
normally, in contrast with the behavior of the ${\cal O}$ structure
discussed above. Finally, let us note that, as shown in
Fig.~\ref{fig14}, the lowest-energy phonons of the ${\cal T}$ phase
are largely dominated by the oxygen cations. This results is in
contrast with our findings for the ${\cal R}$, ${\cal O}$, and ${\cal
  C}$ structures. Such a differentiated behavior is probably related
to the fact that, unlike to all the other phases, the basic building
block of the ${\cal T}$ structure are O$_{5}$ pyramids [see
  Fig.~\ref{fig1}(c)]; having so many oxygen-dominated low-frequency
modes suggests that such pyramids are more easily deformable than the
rather rigid O$_{6}$ octahedra characteristic of the other phases.

\subsection{Pressure-induced transitions at 0~K}
\label{subsec:p-induced}

\begin{figure}
\centerline{ \includegraphics[width=1.00\linewidth]{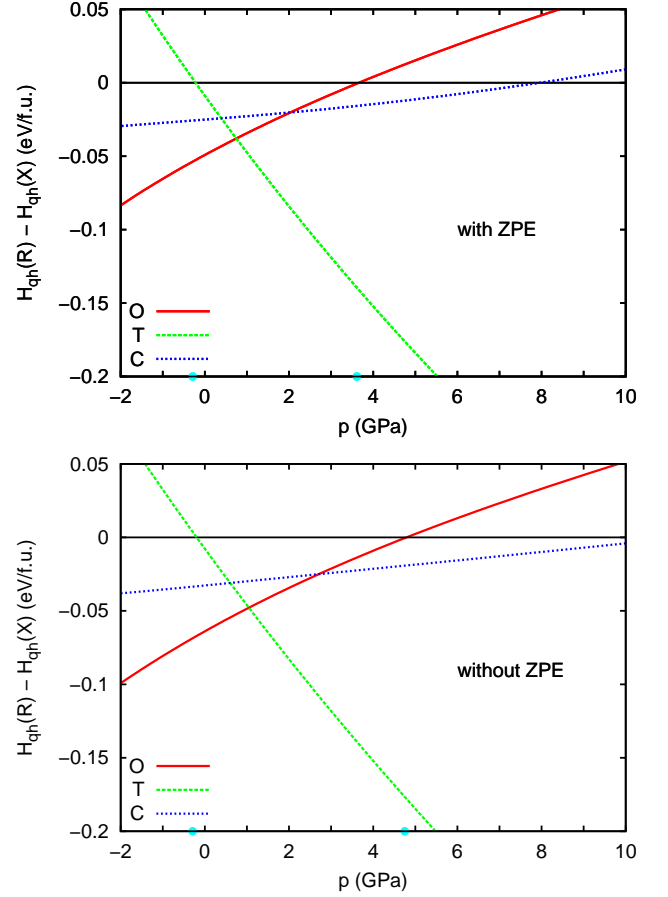}}
\caption{(Color online) Calculated enthalpy energy differences among the four 
  crystal structures analyzed in this work, at $T = 0$~K and expressed as a
  function of $p$. Results obtained when considering
  (resp. neglecting) ZPE corrections are shown in the top
  (resp. bottom) panel. Blue dots in the pressure axis mark the
  occurrence of first-order phase transitions (i.e., $\Delta H_{\rm
    qh} (p_{t}) = 0$).}
\label{fig15}
\end{figure}

\begin{table}
\begin{center}
\label{tab:structure}
\begin{tabular}{c c c c c}
\hline
\hline
$  $ & $  $ & $  $ & $  $ & $ $ \\
\multicolumn{2}{c}{$R3c-G$} & $ a = 5.606$~\AA &  $ b = 5.606$~\AA  &  $ c = 13.950$~\AA  \\
\multicolumn{2}{c}{${\rm (P~=~3.6~GPa)}$} & $ \alpha = 90~^{\circ}$ &  $ \beta = 90~^{\circ}$  &  $ \gamma = 120~^{\circ} $  \\
$  $ & $  $ & $  $ & $  $ & $ $ \\
\hline
$  $ & $  $ & $  $ & $  $ & $ $ \\
${\rm Atom} $ & $ {\rm Wyc.} $ & $ x $ & $ y $ & $ z $ \\
$  $ & $  $ & $  $ & $  $ & $ $ \\
${\rm Bi} $ & $ 6a $ & $ 0.0 $ & $ 0.0 $ & $ 0.4959 $ \\
${\rm Fe} $ & $ 6a $ & $ 0.0 $ & $ 0.0 $ & $ 0.2734 $ \\
${\rm O } $ & $18b $ & $ 0.4186 $ & $-0.0174 $ & $ 0.0402 $ \\
$  $ & $  $ & $  $ & $  $ & $ $ \\
\hline
\hline
$  $ & $  $ & $  $ & $  $ & $ $ \\
\multicolumn{2}{c}{$Pnma-G$} & $ a = 5.696$~\AA &  $ b = 7.838$~\AA  &  $ c = 5.465$~\AA  \\
\multicolumn{2}{c}{${\rm (P~=~3.6~GPa)}$} & $ \alpha = 90~^{\circ}$ &  $ \beta = 90~^{\circ}$  &  $ \gamma = 90~^{\circ} $  \\
$  $ & $  $ & $  $ & $  $ & $ $ \\
\hline
$  $ & $  $ & $  $ & $  $ & $ $ \\
${\rm Atom} $ & $ {\rm Wyc.} $ & $ x $ & $  y $ & $  z $ \\
$  $ & $  $ & $  $ & $  $ & $ $ \\
${\rm Bi} $ & $ 4c $ & $ 0.0512 $ & $ 0.25 $ & $ 0.5098 $ \\
${\rm Fe} $ & $ 4a $ & $ 0.0 $ & $ 0.0 $ & $ 0.0 $ \\
${\rm O } $ & $ 4c $ & $-0.0285 $ & $ 0.25 $ & $ 0.0960$ \\
${\rm O } $ & $ 8d $ & $ 0.1998 $ & $ -0.0469 $ & $ 0.3044$ \\
$  $ & $  $ & $  $ & $  $ & $ $ \\
\hline
\hline
\end{tabular}
\end{center}
\caption{Calculated structural data corresponding to the $p$-induced
  ${\cal R} \to {\cal O}$ transition that is predicted when quantum
  ZPE corrections are considered. Wyckoff positions were generated
  with the ISOTROPY package.\cite{isotropy}}
\end{table}

In this section we analyze the thermodynamic stability of the four
studied crystal structures under hydrostatic pressure at $T = 0$~K. We
take into account ZPE corrections and consider also negative
pressures.

In Fig.~\ref{fig15} we plot the enthalpy energy (i.e., $H = E + pV$)
of the ${\cal O}$, ${\cal T}$, and ${\cal C}$ phases as a function of
$p$, taking the result for the ${\cal R}$ structure as the
pressure-dependent zero of enthalpy. A first-order transformation
between phases $A$ and $B$ occurs at pressure $p_{t}$ when the
enthalpy energy difference $\Delta H (p_{t}) \equiv H_{A} - H_{B}$
becomes zero.  In all the cases we present the results obtained both
when neglecting ZPE corrections (i.e., for $E = E_{\rm eq}$ and $p = -
\partial E_{\rm eq} / \partial V$) and when fully considering them
(i.e., for $E = E_{\rm eq} + F_{\rm harm} ( T \to 0)$ and $p = -
\partial [E_{\rm eq} + F_{\rm harm} ( T \to 0)] / \partial
V$). Additional phonon and static energy calculations were performed
whenever required in order to compute accurate enthalpies in the
pressure interval $-$2~GPa $\le p \le$~10~GPa.

As we increase the pressure, we find two phase transitions of the
${\cal T} \to {\cal R}$ and ${\cal R} \to {\cal O}$ types. The ${\cal
  T} \to {\cal R}$ transition occurs at $-$0.3(1)~GPa and the
associated volume change is $\Delta V = 6.76$~\AA$^{3}$/f.u.; at this
transition pressure, the ${\cal T}$ phase presents a volume of
71.94~\AA$^{3}$/f.u. and a very large $c/a$ ratio of about 2. The
${\cal R} \to {\cal O}$ transition occurs at 3.6(1)~GPa, and the
volume changes from 63.04~\AA$^{3}$/f.u. to 61.13~\AA$^{3}$/f.u.; the
corresponding structural data is given in Table~II.  Interestingly,
the pressure-dependence of the enthalpies shown in Fig.~\ref{fig15}
resemble the results reported in Fig.~\ref{fig12} for $F_{\rm qh}$ as
a function of temperature.  In particular, under compression the
${\cal T}$ phase becomes higher in enthalpy than the rest, and the
enthalpy of the ${\cal O}$ phase turns out to be the smallest. Also,
the ${\cal C}$ phase gets energetically favored over the ${\cal R}$
structure upon increasing pressure, although it never becomes the most
stable structure.

The bottom panel in Fig.~\ref{fig15} shows the enthalpy results
obtained when ZPE corrections are neglected. Interestingly, while the
main trends are conserved, the pressure of the ${\cal R} \to {\cal O}$
transformation turns out to be shifted up to 4.8(1)~GPa. This result
shows that atomic quantum delocalization effects in perovskite oxides
may be important for accurate prediction of $p$-induced phase
transitions.

Our results for the ${\cal R} \to {\cal O}$ transformation are
consistent with those of previous theoretical
studies,\cite{ravindran06,dieguez11} the quantitative differences being
related to the varying DFT flavors employed, consideration of
typically-neglected ZPE corrections, and other technicalities.  As
regards the connection with experiment, it is worth noting that we
predict the ${\cal R} \to {\cal O}$ transition to occur at a pressure
(3.6~GPa) that is rather close to the one at which the ${\cal R}$
phase has been observed to transform into a complex structure by
Guennou {\sl et al.}\cite{guennou11} (i.e., $\sim$~4~GPa at room
temperature).  It is therefore tempting to identify the experimentally
detected complex structure with the family of ${\cal C}$ phases of
which we have investigated a representative case; indeed, verifying a
possible ${\cal R} \to {\cal C} \to {\cal O}$ transition sequence was
one of our motivations to investigate the effects of
pressure. However, our calculations render a direct ${\cal R} \to
{\cal O}$ transition, which suggests that the experimentally observed
complex structures might actually be very long-lived meta-stable
states, as opposed to actual equilibrium phases. On the other hand, as
explained in Section~\ref{subsec:complex}, getting accurate
predictions near transition points at which $F_{\rm qh}({\cal R})
\approx F_{\rm qh}({\cal C}) \approx F_{\rm qh}({\cal O})$ is clearly
a challenging task, and many factors can come into play and affect the
results. Hence, we cannot fully discard the possibility that, under
pressure, the ${\cal R}$ structure transforms into a complex
equilibrium phase.

\subsection{The role of the exchange-correlation energy functional}
\label{subsec:lda}

In previous sections we have highlighted that the differences in the
Helmholtz free energies and enthalpies of the ${\cal R}$, ${\cal O}$,
and ${\cal C}$ phases are calculated to be exceedingly small. In such
conditions, our predictions for the equilibrium phase may depend,
among other factors, on the employed exchange-correlation DFT energy
functional. In this sense, Di\'eguez \emph{et al.} already
found\cite{dieguez11} that, in BFO, energy differences between stable
structures depend strongly on the DFT energy functional used, with
variations in $E_{\rm eq}$ that may be as large as 0.1~eV per formula
unit.

To estimate the magnitude of this type of uncertainties in our $\Delta
F_{\rm qh}$ results computed with a PBE+$U$ functional, we repeated
our QH investigation of temperature-driven transitions -- at constant
volume and frozen-spin conditions -- using a LDA+$U$ scheme. Our
LDA+$U$ results show, in accordance with the presented PBE+$U$ study,
that the orthorhombic ${\cal O}$ phase gets thermodynamically
stabilized over the rest of structures at high temperatures, and that
the ${\cal T}$ phase goes steadily higher in free energy. Further, the
LDA+$U$ results indicate that the ${\cal R} \to {\cal O}$ transition
occurs at approximately 500~K, which is much lower than the
experimental result. Interestingly, most of the the discrepancy
between this LDA+$U$ result and our PBE+$U$ prediction (900~K) can be
traced back to the different equilibrium energies in the 0~K limit,
with the phonon contributions to the free energy playing a secondary
role. Indeed, from the PBE+$U$ calculations we get $E_{\rm eq}({\cal
  O}) - E_{\rm eq}({\cal R})$~=$-0.061$~eV/f.u., while the LDA+$U$
result is $-0.016$~eV/f.u. Obviously, the LDA+$U$ functional brings
the ${\cal R}$ and ${\cal O}$ phases much closer in energy, which
leads to the stabilization of the ${\cal O}$ structure at a much lower
temperature. Additionally, the $F_{\rm qh}$ of the ${\cal C}$ phase
remains always about 50~meV/f.u. higher than that of the ${\cal R}$
phase, the difference being weakly dependent on temperature.

Hence, our calculations confirm that quantitative predictions of
transition temperatures are strongly dependent on the employed DFT
functional. We can also conclude that the LDA functional does not
capture properly the relative stability of the ${\cal R}$ and ${\cal
  O}$ phases of BFO, and that the PBE functional is a much better
choice. In this sense, our work ratifies the conclusions presented in
Ref.~\onlinecite{dieguez11}.

\section{Conclusions}
\label{sec:conclusions}

We have performed a first-principles study of the $p-T$ phase diagram
of bulk multiferroic BFO relying on quasi-harmonic free energy
calculations. We have analyzed the thermodynamic stability of four
different crystal structures that have been observed, or predicted to
exist, at normal and high $p$ or $T$ conditions and/or in thin films
under epitaxial constraints. In order to incorporate the effects of
spin-phonon coupling on the quasi-harmonic calculation of the
Helmholtz free energies, we have developed an approximate and
technically simple scheme that allows us to model states with varying
degrees of spin disorder.

Consistent with observations, we find that the rhombohedral $R3c$
ferroelectric phase (${\cal R}$ phase) is the ground state of the
material at ambient conditions of pressure. Then, an orthorhombic
$Pnma$ structure (${\cal O}$ phase), which is the vibrationally-softest 
of all the considered structures, is found to stabilize upon increasing
$T$ or $p$. More precisely, two first-order
phase transitions of the $R \to O$ type are predicted to occur at the
thermodynamic states [0~GPa, 1300(100)~K] and [3.6(1)~GPa, 0~K].

Additionally, a representative of the so-called {\sl nano-twinned}
structures recently predicted to occur in BFO\cite{prosandeev12} has
been analyzed in this work. This phase is found to display elastic and
vibrational features that are reminiscent of the results obtained for
both the ${\cal R}$ and ${\cal O}$ structures, and to become
energetically more stable than the ${\cal R}$ phase upon raising $p$
and $T$. The entropy and enthalpy of the ${\cal O}$ phase, however,
turn out to be more favorable than those of the studied ${\cal C}$
structure over practically all the investigated $p-T$ intervals, and
as a result we do not observe any direct ${\cal R} \to {\cal C}$ or
${\cal C} \to {\cal O}$ transformation. Nevertheless, our results
cannot be conclusive in this point due to the limitations of the study
(only one specific nano-twinned structure is investigated) and
DFT-related accuracy problems that appear when tackling very small
free-energy differences (i.e., of order $1-10$~meV/f.u.). In fact, our
results seem to support the possibility that some nano-twinned
structures may become stable at the boundaries between ${\cal R}$ and
${\cal O}$ phases in the $p-T$ phase diagram of BFO, or at least exist
as long-lived meta-stable phases that are likely to be accessed
depending on the kinetics of the ${\cal R} \to {\cal O}$
transformation.

Finally, we find that a representative of the so-called {\sl
  super-tetragonal} phases of BFO gets energetically destabilized over
the rest of crystal structures by effect of increasing temperature,
due to the fact that its spectrum of phonon frequencies is globally
the stiffest one. This explains why super-tetragonal structures have
never been observed in bulk BFO, in spite of the fact that their
DFT-predicted energies are very close to those of the ${\cal R}$ and
${\cal O}$ phases. Interestingly, the investigated super-tetragonal
structure is also destabilized upon hydrostatic compression.

As far as we know, our work is the first application of the
quasi-harmonic free energy method to the study of the phase diagram of
a multiferroic perovskite system. The main advantages of this approach
are that is computationally affordable, can be straightforwardly
applied to the study of crystals, and naturally incorporates
zero-point energy corrections. Among its shortcomings, we note that it
can be exclusively applied to the analysis of vibrationally stable
crystal structures; further, it only incorporates anharmonic effects
{\sl via} the volume-dependence of the phonon frequencies and
corresponding treatment of the thermal expansion, which may be a
questionable approximation at high temperatures. Nevertheless, we may
think of several physically interesting (and computationally very
challenging) situations involving BFO-related multiferroics in which
the present approach can prove to be especially useful. A particularly
interesting possibility pertains to the study of solid solutions,
i.e., bulk mixtures of two or more compounds, at finite
temperatures. By assuming simple (or not so simple) relations among
the free energy of the composite system, the relative proportion
between the species, and the vibrational features of the integrating
bulk compounds, one may be able to estimate the phase boundaries in
the complicated $x$-$p$-$T$ phase diagrams at reasonably modest
computational effort. In this regard, the BiFeO$_{3}$-BiCoO$_{3}$ and
BiFeO$_{3}$-LaFeO$_{3}$ solid solutions emerge as particularly
attractive cases, since the application electric fields in suitably
prepared materials can potentially trigger the switching between
different ferroelectric-ferroelectric and ferroelectric-paraelectric
phases.\cite{dieguez11b,otto12}

Beyond possible applications, studying the BiFeO$_{3}$-BiCoO$_{3}$
solid solution is by itself very interesting. On the one hand, this is a case
involving transitions between phases that are very dissimilar
structurally (super-tetragonal and quasi-rhombohedral), and which have
different magnetic orders (C-AFM and G-AFM). Hence, in this case we
can expect spin-phonon effects to have a larger impact in the free
energy, which would allow us to better test the spin-phonon
quasi-harmonic approach that we have introduced in the present
work. Additionally, the treatment of the C-AFM order requires a more
complicated model of exchange interactions, involving at least two
(preferably three\cite{escorihuela12}) coupling constants. Hence,
treating C-AFM phases requires a extension of the scheme here
presented, so that it can easily tackle more general situations. Work
in this direction is already in progress within our group.

\acknowledgments This work was supported by MINECO-Spain [Grants
  No. MAT2010-18113 and No. CSD2007-00041] and the CSIC JAE-doc
program (C.C.). We used the supercomputing facilities provided by RES
and CESGA, and the VESTA software\cite{vesta} for the preparation of some 
figures. The authors acknowledge very stimulating discussions with
Massimiliano Stengel.


\begin{thebibliography}{30}
\bibitem{aclaration-pnma} In Arnold's original work the paraelectric
  $\beta$-phase is designated as $Pbnm$. Actually orthorhombic
  ABO$_{3}$ perovskites have space group $Pnma$ (in accordance with
  crystallographic conventions), and both $Pbnm$ and $Pnma$ space
  groups are equivalent and related via a simple transformation.
  Within the physics community, however, refering to the space group
  $Pbnm$ appears to be more popular.
\bibitem{aclaration-polarity} Given the lattice-periodic part of a
  particular phonon eigenmode $\hat{u}_{s,\kappa\alpha}$, we calculate
  its polarity as $P_{s,\alpha} = \sum_{\kappa\beta}
  Z_{\kappa\beta,\alpha}^{*} \hat{u}_{s,\kappa\beta}$, where $\kappa$
  runs over atoms in the unit cell, $\alpha$ and $\beta$ are Cartesian
  directions, and $\boldsymbol{Z}^{*}_{\kappa}$ is the Born effective
  charge tensor for atom $\kappa$.
\end{thebibliography}
\end{document}